\documentclass[fleqn,usenatbib]{mnras}

\usepackage{newtxtext,newtxmath}
\usepackage[T1]{fontenc}

\DeclareRobustCommand{\VAN}[3]{#2}
\let\VANthebibliography\thebibliography
\def\thebibliography{\DeclareRobustCommand{\VAN}[3]{##3}\VANthebibliography}

\usepackage{graphicx}
\usepackage{subcaption}
\usepackage{amsmath}
\usepackage{bm}
\usepackage{multirow}
\usepackage{acronym}
\usepackage[normalem]{ulem}
\usepackage[dvipsnames]{xcolor}
\usepackage{hyperref}

\DeclareMathOperator{\erf}{erf}

\newcommand{\priorPopAXcLo}{4}  
\newcommand{\priorPopAXcHi}{7}  
\newcommand{\priorPopALamXcLo}{8}  
\newcommand{\priorPopALamXcHi}{45}  
\newcommand{\priorPopAAlphaOneSpinLo}{0.5}  
\newcommand{\priorPopAAlphaOneSpinHi}{5}  
\newcommand{\priorPopAAlphaTwoSpinLo}{1}  
\newcommand{\priorPopAAlphaTwoSpinHi}{15}  
\newcommand{\priorPopABetaOneSpinLo}{0.5}  
\newcommand{\priorPopABetaOneSpinHi}{15}  
\newcommand{\priorPopABetaTwoSpinLo}{0.5}  
\newcommand{\priorPopABetaTwoSpinHi}{15}  
\newcommand{\priorPopAWeightSpinLo}{0.001}  
\newcommand{\priorPopAWeightSpinHi}{0.999}  
\newcommand{\priorPopAAlphaOneEccLo}{1.001}  
\newcommand{\priorPopAAlphaOneEccHi}{50}  
\newcommand{\priorPopAAlphaTwoEccLo}{1.5}  
\newcommand{\priorPopAAlphaTwoEccHi}{8}  
\newcommand{\priorPopABetaOneEccLo}{3}  
\newcommand{\priorPopABetaOneEccHi}{45}  
\newcommand{\priorPopABetaTwoEccLo}{2}  
\newcommand{\priorPopABetaTwoEccHi}{15}  
\newcommand{\priorPopAWeightEccLo}{0.001}  
\newcommand{\priorPopAWeightEccHi}{0.999}  
\newcommand{\priorPopAKappaZLo}{0}  
\newcommand{\priorPopAKappaZHi}{6}  
\newcommand{\priorPopBLamMLo}{\ensuremath{-3}}  
\newcommand{\priorPopBLamMHi}{4}  
\newcommand{\priorPopBMuOneSpinLo}{0.05}  
\newcommand{\priorPopBMuOneSpinHi}{0.45}  
\newcommand{\priorPopBSigmaOneSpinLo}{0.001}  
\newcommand{\priorPopBSigmaOneSpinHi}{0.3}  
\newcommand{\priorPopBMuTwoSpinLo}{0.45}  
\newcommand{\priorPopBMuTwoSpinHi}{0.99}  
\newcommand{\priorPopBSigmaTwoSpinLo}{0.001}  
\newcommand{\priorPopBSigmaTwoSpinHi}{0.3}  
\newcommand{\priorPopBWeightSpinLo}{0.001}  
\newcommand{\priorPopBWeightSpinHi}{0.999}  
\newcommand{\priorPopBMuOneEccLo}{0.05}  
\newcommand{\priorPopBMuOneEccHi}{0.3}  
\newcommand{\priorPopBSigmaOneEccLo}{0.01}  
\newcommand{\priorPopBSigmaOneEccHi}{0.2}  
\newcommand{\priorPopBMuTwoEccLo}{0.3}  
\newcommand{\priorPopBMuTwoEccHi}{0.6}  
\newcommand{\priorPopBSigmaTwoEccLo}{0.02}  
\newcommand{\priorPopBSigmaTwoEccHi}{0.3}  
\newcommand{\priorPopBWeightEccLo}{0.001}  
\newcommand{\priorPopBWeightEccHi}{0.999}  
\newcommand{\priorPopBKappaZLo}{0}  
\newcommand{\priorPopBKappaZHi}{6}  

\newcommand{\aggPopAXcEmriRelPctMed}{41}  
\newcommand{\aggPopAXcEmriWidMin}{0.51}  
\newcommand{\aggPopAXcEmriWidMax}{1.4}  
\newcommand{\aggPopALamXcEmriRelPctMed}{77}  
\newcommand{\aggPopAKappaZEmriRelPctMed}{24}  
\newcommand{\aggPopAKappaZEmriWidMin}{1.2}  
\newcommand{\aggPopAKappaZEmriWidMax}{2.2}  
\newcommand{\aggPopAXcZtfWidMin}{0.96}  
\newcommand{\aggPopAXcZtfWidMax}{1.5}  
\newcommand{\aggPopAKappaZZtfWidMin}{0.6}  
\newcommand{\aggPopAKappaZZtfWidMax}{2.1}  
\newcommand{\aggPopAXcLsstWidMed}{0.31}  
\newcommand{\aggPopALamXcLsstWidMed}{16}  
\newcommand{\aggPopAKappaZLsstWidMin}{0.095}  
\newcommand{\aggPopAKappaZLsstWidMed}{0.11}  
\newcommand{\aggPopAKappaZLsstWidMax}{0.14}  
\newcommand{\aggPopAXcEmriLsstWidMin}{0.13}  
\newcommand{\aggPopAXcEmriLsstWidMax}{0.65}  
\newcommand{\aggPopBLamMEmriRelPctMed}{43}  
\newcommand{\aggPopBLamMLsstWidMed}{0.44}  
\newcommand{\aggPopBKappaZLsstWidMin}{0.1}  
\newcommand{\aggPopBKappaZLsstWidMax}{0.14}  
\newcommand{\aggPopAXcEmriZtfGain}{1.1}  
\newcommand{\aggPopAXcEmriLsstGain}{4}  
\newcommand{\aggPopALamXcEmriZtfGain}{1.1}  
\newcommand{\aggPopALamXcEmriLsstGain}{1.9}  
\newcommand{\aggPopAKappaZEmriZtfGain}{2}  
\newcommand{\aggPopAKappaZEmriLsstGain}{13}  
\newcommand{\aggPopBLamMEmriZtfGain}{1.3}  
\newcommand{\aggPopBLamMEmriLsstGain}{7}  
\newcommand{\aggPopBKappaZEmriZtfGain}{2}  
\newcommand{\aggPopBKappaZEmriLsstGain}{13}  
\newcommand{\aggPopAEmriFracZtfPctMin}{1.2}  
\newcommand{\aggPopAEmriFracZtfPctMed}{3.4}  
\newcommand{\aggPopAEmriFracZtfPctMax}{14}  
\newcommand{\aggPopATdePerEmriZtfMin}{2.7}  
\newcommand{\aggPopATdePerEmriZtfMax}{84.3}  
\newcommand{\aggPopAEmriFracLsstPctMax}{0.11}  
\newcommand{\aggPopBLamMSignAgreeZtfPctMed}{90}  
\newcommand{\aggPopBLamMSignAgreeEmriPctMed}{96}  
\newcommand{\aggBothInfoEmriMin}{4.4}  
\newcommand{\aggBothInfoEmriMax}{8.8}  
\newcommand{\aggBothInfoZtfMin}{1.5}  
\newcommand{\aggBothInfoZtfMax}{3.4}  
\newcommand{\aggBothInfoLsstMin}{5.8}  
\newcommand{\aggBothInfoLsstMax}{7.5}  
\newcommand{\aggBothInfoEmriZtfMinusEmriMin}{0.15}  
\newcommand{\aggBothInfoEmriZtfMinusEmriMax}{1.8}  
\newcommand{\aggBothInfoEmriLsstMinusEmriMin}{4.1}  
\newcommand{\aggBothInfoEmriLsstMinusEmriMax}{5.3}  
\newcommand{\aggBothInfoEmriMinusTdeMin}{3}  
\newcommand{\aggBothInfoEmriMinusTdeMax}{6.8}  
\newcommand{\aggBothJointShiftStdZtfMax}{3.3}  
\newcommand{\aggBothJointOverTdeWidthZtfMaxDevPct}{69}  
\newcommand{\aggBothJointOverTdeWidthLsstMin}{0.76}  
\newcommand{\aggBothJointOverTdeWidthLsstMax}{1.1}  
\newcommand{\aggBothJointShiftStdLsstMax}{0.94}  
\newcommand{\aggBothZtfOnLsstMaxChangePct}{11}  

\newcommand{\simNumRealisations}{100}  
\newcommand{\simEmriPerTdeZtfMin}{0.021}  
\newcommand{\simEmriPerTdeZtfMax}{0.043}  
\newcommand{\simEmriPerTdeLsstMinTimesTenThousand}{4.6}  
\newcommand{\simEmriPerTdeLsstMaxTimesTenThousand}{6.8}  
\newcommand{\simObservedZtfTde}{38}  

\newcommand{\tabCountEmriStarHU}{\ensuremath{5^{+3}_{-3}}}  
\newcommand{\tabCountZtfStarHU}{\ensuremath{205^{+118}_{-96}}}  
\newcommand{\tabCountLsstStarHU}{\ensuremath{7389^{+1065}_{-860}}}  
\newcommand{\tabUstarEmriStarHU}{9}  
\newcommand{\tabUstarZtfStarHU}{213}  
\newcommand{\tabUstarLsstStarHU}{8074}  
\newcommand{\tabRatioZtfStarHU}{\ensuremath{2.2^{+2.7}_{-1.3}}}  
\newcommand{\tabRatioLsstStarHU}{\ensuremath{6.1^{+5.0}_{-3.5}}}  
\newcommand{\tabCountEmriSigHU}{\ensuremath{5^{+3}_{-3}}}  
\newcommand{\tabCountZtfSigHU}{\ensuremath{117^{+98}_{-86}}}  
\newcommand{\tabCountLsstSigHU}{\ensuremath{8356^{+727}_{-853}}}  
\newcommand{\tabUstarEmriSigHU}{7}  
\newcommand{\tabUstarZtfSigHU}{19}  
\newcommand{\tabUstarLsstSigHU}{6965}  
\newcommand{\tabRatioZtfSigHU}{\ensuremath{4.4^{+11.3}_{-3.2}}}  
\newcommand{\tabRatioLsstSigHU}{\ensuremath{5.8^{+4.6}_{-3.5}}}  
\newcommand{\tabCountEmriStarNU}{\ensuremath{3^{+5}_{-2}}}  
\newcommand{\tabCountZtfStarNU}{\ensuremath{144^{+114}_{-75}}}  
\newcommand{\tabCountLsstStarNU}{\ensuremath{5299^{+728}_{-783}}}  
\newcommand{\tabUstarEmriStarNU}{3}  
\newcommand{\tabUstarZtfStarNU}{253}  
\newcommand{\tabUstarLsstStarNU}{5656}  
\newcommand{\tabRatioZtfStarNU}{\ensuremath{2.1^{+6.0}_{-1.7}}}  
\newcommand{\tabRatioLsstStarNU}{\ensuremath{6.2^{+9.8}_{-4.6}}}  
\newcommand{\tabCountEmriSigNU}{\ensuremath{3^{+4}_{-2}}}  
\newcommand{\tabCountZtfSigNU}{\ensuremath{75^{+107}_{-48}}}  
\newcommand{\tabCountLsstSigNU}{\ensuremath{5938^{+694}_{-838}}}  
\newcommand{\tabUstarEmriSigNU}{4}  
\newcommand{\tabUstarZtfSigNU}{53}  
\newcommand{\tabUstarLsstSigNU}{5604}  
\newcommand{\tabRatioZtfSigNU}{\ensuremath{4.3^{+10.7}_{-3.5}}}  
\newcommand{\tabRatioLsstSigNU}{\ensuremath{6.0^{+5.1}_{-4.5}}}  
\newcommand{\tabCountEmriStarHV}{\ensuremath{3^{+4}_{-2}}}  
\newcommand{\tabCountZtfStarHV}{\ensuremath{132^{+125}_{-79}}}  
\newcommand{\tabCountLsstStarHV}{\ensuremath{5252^{+678}_{-719}}}  
\newcommand{\tabUstarEmriStarHV}{3}  
\newcommand{\tabUstarZtfStarHV}{164}  
\newcommand{\tabUstarLsstStarHV}{5409}  
\newcommand{\tabRatioZtfStarHV}{\ensuremath{2.3^{+5.8}_{-1.9}}}  
\newcommand{\tabRatioLsstStarHV}{\ensuremath{5.9^{+6.8}_{-4.3}}}  
\newcommand{\tabCountEmriSigHV}{\ensuremath{3^{+4}_{-3}}}  
\newcommand{\tabCountZtfSigHV}{\ensuremath{76^{+103}_{-49}}}  
\newcommand{\tabCountLsstSigHV}{\ensuremath{6062^{+727}_{-740}}}  
\newcommand{\tabUstarEmriSigHV}{7}  
\newcommand{\tabUstarZtfSigHV}{53}  
\newcommand{\tabUstarLsstSigHV}{6434}  
\newcommand{\tabRatioZtfSigHV}{\ensuremath{3.8^{+12.0}_{-3.8}}}  
\newcommand{\tabRatioLsstSigHV}{\ensuremath{5.0^{+6.0}_{-5.0}}}  
\newcommand{\tabCountEmriStarNV}{\ensuremath{2^{+3}_{-2}}}  
\newcommand{\tabCountZtfStarNV}{\ensuremath{96^{+80}_{-66}}}  
\newcommand{\tabCountLsstStarNV}{\ensuremath{3721^{+810}_{-619}}}  
\newcommand{\tabUstarEmriStarNV}{3}  
\newcommand{\tabUstarZtfStarNV}{106}  
\newcommand{\tabUstarLsstStarNV}{3678}  
\newcommand{\tabRatioZtfStarNV}{\ensuremath{2.3^{+5.8}_{-2.3}}}  
\newcommand{\tabRatioLsstStarNV}{\ensuremath{5.8^{+7.6}_{-5.8}}}  
\newcommand{\tabCountEmriSigNV}{\ensuremath{2^{+4}_{-2}}}  
\newcommand{\tabCountZtfSigNV}{\ensuremath{63^{+79}_{-44}}}  
\newcommand{\tabCountLsstSigNV}{\ensuremath{4355^{+651}_{-689}}}  
\newcommand{\tabUstarEmriSigNV}{3}  
\newcommand{\tabUstarZtfSigNV}{145}  
\newcommand{\tabUstarLsstSigNV}{4953}  
\newcommand{\tabRatioZtfSigNV}{\ensuremath{3.0^{+12.6}_{-3.0}}}  
\newcommand{\tabRatioLsstSigNV}{\ensuremath{4.6^{+8.8}_{-4.6}}}  

\title[Constraining the MBH population with EMRIs and TDEs]{Disrupted stars and captured compact objects: Combining observations from multiple messengers
to infer the population of massive black holes}

\author[S.
Singh et al.]{
S. Singh,$^{1}$\thanks{E-mail: s.singh.3@research.gla.ac.uk}
L. Broggi,$^{2,3}$
C. P. L. Berry$^{1}$
and M. Volonteri$^{4}$
\\
$^{1}$SUPA, School of Physics and Astronomy, University of Glasgow, Glasgow G12 8QQ, UK\\
$^{2}$Dipartimento di Fisica ``G. Occhialini", Università degli Studi di Milano-Bicocca, Piazza della Scienza 3, 20126 Milano, Italy\\
$^{3}$INFN, Sezione di Milano-Bicocca, Piazza della Scienza 3, 20126 Milano, Italy\\
$^{4}$Sorbonne Université, CNRS, UMR7095, Institut d’Astrophysique de Paris, 98bis boulevard Arago, F-75014, Paris, France
}

\pubyear{\the\year{}}

\begin{document}
\label{firstpage}
\pagerange{\pageref{firstpage}--\pageref{lastpage}}
\maketitle

\acrodef{EMRI}[EMRI]{extreme mass-ratio inspiral}
\acrodef{TDE}[TDE]{tidal disruption event}
\acrodef{CO}[CO]{compact object}
\acrodef{LISA}[\textit{LISA}]{\textit{Laser Interferometer Space Antenna}}
\acrodef{ZTF}[ZTF]{Zwicky Transient Facility}
\acrodef{LSST}[LSST]{Legacy Survey of Space and Time}
\acrodef{GW}[GW]{gravitational wave}
\acrodef{SNR}[SNR]{signal-to-noise ratio}
\acrodef{MBH}[MBH]{massive black hole}
\acrodef{sBH}[sBH]{stellar-mass black hole}
\acrodef{AGN}[AGN]{active galactic nucleus}
\acrodefplural{AGN}[AGNs]{active galactic nuclei}
\acrodef{NSC}[NSC]{nuclear stellar cluster}
\acrodef{MLP}[MLP]{\emph{multi-layer perceptron}}
\acrodef{CI}[CI]{credible interval}
\acrodef{ML}[ML]{machine-learning}
\acrodef{FP}[FP]{Fokker--Planck}
\acrodef{GSMF}[GSMF]{galaxy stellar mass function}
\acrodef{MBHMF}[MBHMF]{\ac{MBH} mass function}
\acrodef{LMM}[LMM]{last major merger}
\acrodef{CDF}[CDF]{cumulative distribution function}
\acrodef{PPD}[PPD]{posterior predictive distribution}
\acrodef{KS}[KS]{Kolmogorov--Smirnov}

\begin{abstract}
Massive black holes (MBHs) are found in galactic centres across the Universe, but the mechanisms governing their formation and evolution remain elusive.
Tidal disruption events (TDEs) and extreme mass-ratio inspirals (EMRIs) both arise from stellar dynamics in the nuclear star clusters around MBHs, and are observed with electromagnetic surveys and with gravitational waves, respectively.
Overcoming the scarcity of coincident detections, we formulate how to combine gravitational-wave detections from \acs{LISA}-EMRIs with observations from ZTF-TDEs to measure properties of the MBH population.
Building on two-dimensional Fokker--Planck simulations of EMRIs and TDEs, we model how their populations are affected by the MBH--host scaling relation, the nucleation fraction and the MBH occupation fraction, and analyse the resulting catalogues using joint hierarchical inference accounting for their individual selection biases.
With current ZTF-TDE detections, we can already identify combinations of MBH formation that are more favoured and consistent with the observed TDE population, and derive corresponding predictions for the EMRI population detectable by \acs{LISA}.
We find that across a range of different astrophysical assumptions, the ratio of detectable TDEs to EMRIs is high, with ZTF detecting $\aggPopATdePerEmriZtfMin$--$\aggPopATdePerEmriZtfMax$ times as many TDEs as \acs{LISA} detects EMRIs. 
TDEs, through their numbers, set the shared mass and redshift distributions: LSST constrains the characteristic MBH mass to a $90\%$ credible width of $\aggPopAXcLsstWidMed~\mathrm{dex}$ and the power-law index of the redshift evolution to a precision of $\aggPopAKappaZLsstWidMed$.
EMRIs make up $\sim \aggPopAEmriFracZtfPctMed\%$ of the joint catalogue, yet carry more information than the whole ZTF sample and are the only probe of MBH spin and orbital eccentricity.
TDEs will therefore constrain the abundance of low-mass MBHs before \acs{LISA} flies, while EMRIs will measure properties that electromagnetic surveys cannot.
EMRI and TDE observations hence offer complementary insights into MBH growth and evolution that neither can provide alone.
\end{abstract}



\section{Introduction}
\label{sec:introduction}
\Acp{MBH} residing in the centre of galaxies are abundant in our Universe.
Some of the \acp{MBH} hosted in low-to-intermediate galaxies ($\sim10^8$--$10^{11}~M_\odot$) are surrounded by \acp{NSC} with densities reaching $10^6$--$10^7~M_\odot~\mathrm{pc^{-3}}$~\citep{Carollo:1997xvt, Neumayer:2020gno, Pechetti:2019nmt}.
These \acp{NSC} provide a reservoir of \acp{CO} such as \acp{sBH}, white dwarfs, neutron and regular stars~\citep{Graham:2009un, Merritt:2009mr, Georgiev:2016jfa, Neumayer:2020gno}.
The dynamical evolution of stars and \acp{CO} in the \ac{NSC} is driven by the gravitational potential of the \ac{MBH} and its environment ~\citep{2013degn.book.....M}.
Interactions between a \ac{MBH} and the components of its \ac{NSC} give rise to the possibility of observing \aclp{EMRI}~\citep[\acsp{EMRI}; \acused{EMRI}][]{Amaro-Seoane:2012lgq, Berry:2019wgg} and \aclp{TDE}~\citep[\acsp{TDE}; \acused{TDE}][]{Gezari:2021bmb}.

Within the \ac{NSC}, regular stars and white dwarfs undergoing close encounters with the central \ac{MBH} can be completely or partially disrupted by tidal forces.
The nature of the encounter depends on the mass of the \ac{MBH} and the proximity of the interaction, and a star is tidally disrupted only if the tidal radius lies outside the event horizon.
This restricts white dwarf disruptions to \ac{MBH} masses of $\lesssim10^5~M_\odot$~\citep{Sesana:2008zc} and for regular stars to $\lesssim10^7~M_\odot$~\citep{Lodato:2010xs}; above these masses, stars plunge directly into the \ac{MBH}.
Within this regime, the disruption can be partial or full depending on the penetration factor.
These disruptions lead to luminous transients with typical light curves~\citep{Lodato:2008fr} that can exceed the \ac{MBH} Eddington luminosity by orders of magnitude~\citep{Lodato:2010xs}.
Multiple \acp{TDE} have been observed~\citep{vanVelzen:2020cwu, Saxton:2021oaw, Yao:2023rbr, Hammerstein:2022wia} so far in optical~\citep{Bellm:2018cps}, ultraviolet and X-ray~\citep[][]{Saxton:2021oaw}, and these detections are expected to grow with upcoming observatories like \ac{LSST}/Vera Rubin~\citep{Karmen:2026ixp}.

Other \acp{CO} in the \ac{NSC} (especially \acp{sBH}) will not be tidally disrupted, leading to two possible outcomes.
Most of the \acp{sBH} in close proximity to the \ac{MBH} will undergo direct plunges into the \ac{MBH}.
A small fraction of \acp{CO} can instead become gravitationally bound to the \ac{MBH}, forming a highly asymmetric-mass binary effectively decoupled from the rest of the \ac{NSC}~\citep{Amaro-Seoane:2012lgq,LISA:2022yao}.
Such a binary system is known as an \ac{EMRI}.
As the \ac{CO} gradually loses energy and angular momentum through \acp{GW} emission~\citep{Peters:1964zz,Amaro-Seoane:2007osp}, the orbit shrinks slowly.
Due to this mass ratio of the binary, an \ac{EMRI} can spend $\sim10^5$--$10^6$ orbital cycles in the millihertz \ac{GW} band ($\sim10^{-4}$--$10^{-1}~\mathrm{Hz}$); this makes them a target for the \ac{LISA} mission~\citep[][]{LISA:2024hlh}.

\acp{TDE} are already an established transient class: wide-field, high-cadence optical surveys have moved the field from a handful of X-ray-selected candidates to samples of tens of spectroscopically classified events with well-sampled light curves~\citep{Hammerstein:2022wia, vanVelzen:2020cwu, Yao:2023rbr}.
The observed \ac{TDE} light curve is not a direct probe of the disruption itself, but of how the bound debris circularises, accretes, and reprocesses the released energy.
The peak luminosity, rise time and decay rate depend on the \ac{MBH} mass, stellar mass, depth of the encounter, and on the efficiency with which accreted rest-mass energy is radiated~\citep{Gezari:2021bmb, Mockler:2018xne}.
The number of detected sources~\citep{Bricman:2019mcg, Karmen:2026ixp, LSST:2008ijt} and their ability to probe the mass of the \acp{MBH} make them a crucial component of the population analysis \citep{Stone:2014wxa}.

\acp{EMRI}, by contrast, will only be detected following the launch of the \ac{LISA} mission~\citep[currently expected in the mid-2030s;][]{LISA:2024hlh}.
Detection estimates span several orders of magnitude, giving the uncertainty in \ac{NSC} dynamics and \ac{MBH} merger history~\citep{Babak:2017tow,LISA:2022yao}. 
\acp{EMRI} are nevertheless a gold mine for \ac{MBH} information. 
With numerous \ac{GW} cycles in the \ac{LISA} band, they provide measurements of the \ac{MBH} spacetime and \ac{CO} orbit~\citep{Barack:2003fp,Babak:2017tow,Speri:2022upm}. 
These observations in turn can be used to constrain the \ac{MBH} population~\citep{Gair:2004iv,Chapman-Bird:2022tvu}; even a few tens of detections could distinguish between different formation channels and uncover \ac{MBH} environments~\citep{Speri:2022upm,Dyson:2025dlj,Singh:2026sad}. 

Hence, \ac{TDE} observations can potentially constrain \ac{EMRI} predictions in advance of \ac{LISA}, and \ac{EMRI} observations can potentially provide more precise measurements of the \ac{MBH} population than \acp{TDE} alone. 
To combine these distinct populations of observations to measure the properties of \acp{MBH}, we can use hierarchical Bayesian inference.
Hierarchical inference is an established tool for population analyses of compact-binary coalescences, where the shape of the mass, spin and redshift distributions is inferred from a catalogue of individually uncertain events while simultaneously correcting for selection effects~\citep{Mandel:2018mve, Vitale:2020aaz, LIGOScientific:2026ctl}. It has been previously applied to \ac{EMRI} populations specifically~\citep{Chapman-Bird:2022tvu, Singh:2026sad}.
\ac{TDE} demography has historically been treated separately, through volumetric rates and luminosity functions derived from flux-limited samples~\citep{vanVelzen:2020cwu, Hammerstein:2022wia, Yao:2023rbr}.
Because the observed signals from \acp{EMRI} and \acp{TDE} encode complementary information about the \ac{MBH} population, analysing them jointly provides insights inaccessible to either channel alone.

In this work, we build an end-to-end multi-messenger population-inference framework that treats \acp{EMRI} and \acp{TDE} as complementary views of a single \ac{MBH} population.
We construct a simulated galaxy catalogue (Sec.~\ref{sec:astrophysical_setup}) from a cosmologically calibrated \ac{GSMF} populating it with \acp{NSC} using scaling relations and nucleation, and seeding \acp{MBH} using occupation probability.
The number of \ac{EMRI} and \ac{TDE} per galaxy then follow \ac{FP} rates of \citet{Broggi:2022udp}.
We then simulate the responses of \ac{LISA}, \ac{ZTF}, and \ac{LSST}, emulate selection functions, and build a joint hierarchical inference framework to analyse the resulting catalogues (Sec.~\ref{sec:Joint_Hierarchical_Inference}).
The framework builds on \citet{Chapman-Bird:2022tvu, Singh:2025pzh},  extending previous work to \acp{TDE} for the joint population inference of the two messengers under the assumption of a shared \ac{MBH} population.
Our inference results (Sec.~\ref{sec:population_inference_results}) demonstrate that constraints on the underlying population tighten as more observations are included, and we present our conclusions in Sec.~\ref{sec:discussion_conclusion}.

\section{Astrophysical modelling}
\label{sec:astrophysical_setup}

In this section, we describe the astrophysical framework underlying our simulated \ac{EMRI} and \ac{TDE} populations.
Our approach connects the large‑scale evolution of galaxies to the small‑scale dynamics of \acp{NSC} hosting \acp{MBH}.
We begin by specifying the cosmological and galactic initial conditions used to seed a population of nucleated galaxies (Sec.~\ref{subsec:galactic_properties}).
We then describe how these galaxies evolve and form dense \acp{NSC} around their central \acp{MBH}.
The structure of these clusters is not static. Gravitational relaxation, which drives mass segregation, continuously reshapes the nuclear stellar distribution, while major galaxy and \ac{MBH} mergers can further alter the central stellar cusp. 
Major \ac{MBH} mergers can partially or entirely disrupt the cusp, after which it regrows over a relaxation timescale, while galaxy mergers can also help rebuild it through merger-induced nuclear starbursts (Sec.~\ref{subsec:NSCevolution_cuspregrowth}).
This time-dependent evolution, together with the \ac{MBH} mass, directly sets the \ac{EMRI} and \ac{TDE} rates for each galaxy (Sec.~\ref{subsec:simulating_EMRIs_TDEs}).
We assign the intrinsic parameters of the sources to specify the signals generation (Sec.~\ref{subsec:EMRI_TDE_properties}).
Finally, we generate simulated universes and catalogue the number of detected \acp{EMRI} and \acp{TDE} (Sec.~\ref{sec:simulated_catalogue}).

\subsection{Galaxy properties}
\label{subsec:galactic_properties}

\acp{EMRI} and \acp{TDE} originate in dense galactic nuclei, where relaxation processes shape the stellar populations that give rise to these events~\citep{Amaro-Seoane:2007osp, Merritt:2005yt, Hopman:2006xn}. 
Since both sources are intrinsically rare, we require a large galaxy sample to seed. 
Hence, we construct a simulated galaxy catalogue over the redshift range $z=[10^{-2}, 8]$ and galaxy masses from $10^8$--$10^{12}~M_\odot$~\citep[cf.][]{Babak:2017tow,LISA:2024hlh}.

\begin{figure}
    \centering
    \includegraphics[width=\columnwidth]{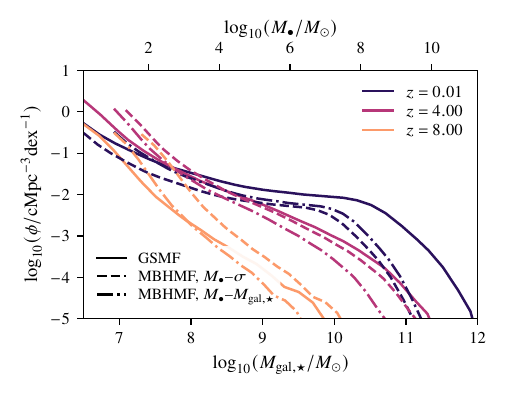}
    \caption{
            Comparison of galaxy and black hole mass functions at different redshifts.
            Solid lines show the galaxy stellar mass function (GSMF; $\phi(M_\mathrm{gal, \star},z)$), while dashed lines indicate the massive black hole mass function (MBHMF; $\phi(M_\bullet)\propto\phi(M_\mathrm{gal, \star},z)$).
            The MBHMF is calculated for two scaling relations.
            The  lower axis shows $\log_{10}(M_{\mathrm{gal}}/M_\odot)$, and the upper axis shows $\log_{10}(M_\bullet/M_\odot)$, for direct comparison.
            The vertical axis represents the logarithm of the comoving number density, $\log_{10}(\phi/\mathrm {cMpc}^{-3}~\mathrm{dex}^{-1})$.
            }
    \label{fig:mass_functions}
\end{figure}

Galaxy positions are drawn assuming isotropy and homogeneity in comoving space.
The redshift range $z=[10^{-2},8]$ is mapped to the comoving distance interval $[\mathcal{R}_\mathrm{min},\mathcal{R}_\mathrm{max}]$ using a flat $\Lambda$--CDM cosmology (we use $H_0 = 70~\mathrm{km\,s^{-1}\,Mpc^{-1}}$ and $\{\Omega_\Lambda,\Omega_\mathrm{m}\} = \{0.7, 0.3\}$).
This is then used to define a spherical shell of comoving volume $\mathcal{V}_\mathrm{c} \propto \mathcal{R}_\mathrm{max}^3-\mathcal{R}_\mathrm{min}^3$.
Sources are then distributed uniformly in this shell based on the number density of the \ac{GSMF}.
The sky coordinates are assigned uniformly and independently of galaxy mass and redshift.
We use the \ac{GSMF} from the \texttt{COLIBRE} simulations~\citep{Chaikin:2026rku}, which has been shown to be in good agreement with observations over the redshift range $z=(0$--$8]$.
In Fig.~\ref{fig:mass_functions} we show a range of \acp{GSMF} $\phi(M_\mathrm{gal, \star},z)$ and the corresponding \acp{MBHMF} $\phi(M_\bullet,z)$, obtained from the \ac{GSMF} through the adopted $M_\bullet$ scaling relation (Sec.~\ref{subsec:galactic_properties}).

Since the production of \acp{EMRI} and \acp{TDE} is expected to be strongly suppressed in non-nucleated systems~\citep{Merritt:2006bm, Antonini:2011tu, Pfister:2020cmc}, we focus on the subset of nucleated galaxies capable of hosting an \ac{NSC} and neglect non-nucleated systems in our modelling.
Neglecting non-nucleated galaxies is expected to have only a small effect on our predicted \ac{EMRI} population, because the \acp{EMRI} we consider (\emph{dry} \acp{EMRI}, formed by two-body relaxation) require the sustained loss-cone refilling of a relaxed stellar cusp (Sec.~\ref{subsec:simulating_EMRIs_TDEs}). 
Non-nucleated galaxies, lacking a dense, relaxed cusp, are not expected to sustain this refilling, and hence do not contribute appreciably to the \ac{EMRI} production.
For \acp{TDE}, however, this remains a conservative choice.
Some off-nuclear (wandering) \acp{MBH} can still produce detectable stellar disruptions~\citep{Gezari:2021bmb, Angus:2022uid}. 
As these systems are not included in our model, our predicted \ac{TDE} rates should be regarded as a lower bound on the total \ac{TDE} population.
In particular, some off-centre \acp{TDE} may arise from stripped satellite nuclei that could also support \ac{EMRI} formation.

To determine whether a given galaxy is nucleated, for a galaxy of mass $M_\mathrm{gal,\star}$ at redshift $z_\mathrm{gal}$, we use the nucleation fractions $f_{\mathrm{nucl}}(M_\mathrm{gal,\star},z)$ from \cite{Hannah:2024dmr, Neumayer:2020gno}.
For each galaxy, we draw a uniform random number $f\in[0,1]$, and the galaxy is seeded with an \ac{NSC} if $f < f_{\mathrm{nucl}}(M_\mathrm{gal,\star},z)$.

Nucleation alone does not guarantee the presence of a central \ac{MBH}.
Not all low-mass galaxies are expected to host \acp{MBH}~\citep{Volonteri:2007ax}, and the resulting \ac{MBH} occupation fraction $f_\mathrm{occ}$ affects predicted \ac{EMRI} and \ac{TDE} sources~\citep{Stone:2014wxa}.
The probability of a galaxy of mass $M_\mathrm{gal,\star}$ hosting an \ac{MBH}-bearing \ac{NSC} is the product of the nucleation fraction and the occupation fraction, $f_\mathrm{nucl}(M_\mathrm{gal,\star}) \times f_\mathrm{occ}(M_\mathrm{gal,\star})$. 
We neglect any correlation between \ac{NSC} nucleation and \ac{MBH} occupation, which remains poorly constrained observationally.

We consider two occupation probabilities: (i) a uniform occupation probability, $f_\mathrm{occ}=1$, where every galaxy hosts an \ac{MBH}, and (ii) the mass-dependent occupation fraction of \citet{Zou:2025mdy} calibrated against $\sim20~\mathrm{yr}$ of \textit{Chandra} X-ray imaging of $1,606$ galaxies within $50~\mathrm{Mpc}$.
The occupation fraction is parametrised as a function of galaxy stellar mass $M_\mathrm{gal,\star}$ as
    \begin{equation}
    \label{eq:focc}
        f_\mathrm{occ}(M_\mathrm{gal,\star}) = \left\{1 + \delta\, \exp\left[\theta\delta(\log_{10} M_\mathrm{gal,\star} - \log_{10} M_0)\right]\right\}^{-1/\delta},
    \end{equation}
with inferred parameters $\log_{10} M_0 = 9.46^{+0.32}_{-0.51}$, $\delta = 3.91^{+4.66}_{-2.53}$, and $\theta = 0.80^{+0.55}_{-0.30}$~\citep[][Table~2 and Fig.~6]{Zou:2025mdy}.

\begin{figure}
    \centering
    \includegraphics[width=\columnwidth]{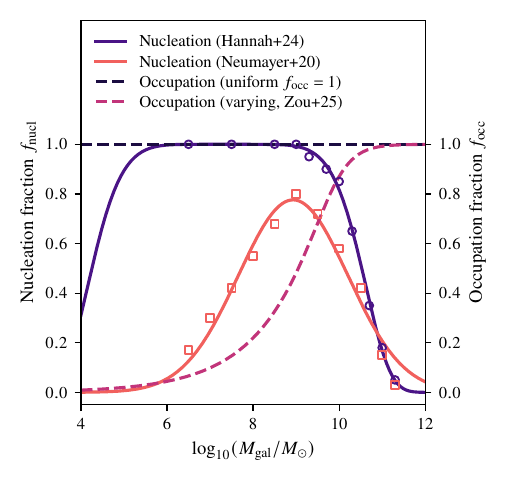}
    \caption{Nucleation fraction $f_{\mathrm{nucl}}(M_\mathrm{gal,\star})$ (left-hand axis), and \ac{MBH} occupation fraction $f_\mathrm{occ}(M_\mathrm{gal,\star})$ (right-hand axis), as a function of logarithmic galaxy stellar mass $\log_{10}(M_{\mathrm{gal},\star})$.
    Solid curves show the two nucleation-fraction prescriptions and the dashed curves show the two occupation-fraction assumptions.
    The combination of the two nucleation-fraction and two occupation-fraction results in the four population scenarios (Table~\ref{tab:simulated_catalogue_scenario_summary}).}
    \label{fig:nucleation_occupation_fraction}
\end{figure}

The combined effect of nucleation and occupation on the number of galaxies relevant for our analysis is shown in Fig.~\ref{fig:nucleation_occupation_fraction}.
Under the \citet{Neumayer:2020gno} prescription, $f_\mathrm{nucl}$ is itself already low ($\lesssim0.2$) at $M_\mathrm{gal,\star}\sim10^4~M_\odot$, so the product is suppressed by nucleation regardless of which occupation model is used, and switching from uniform to varying occupation changes the low-mass population only modestly.
Under the \citet{Hannah:2024dmr} prescription, in contrast, $f_\mathrm{nucl}\approx1$ down to the lowest galaxy masses sampled, so the product is set almost entirely by the occupation fraction at low masses.
Adopting a varying \ac{MBH} occupation model in place of uniform suppresses the low-mass \ac{MBH}-hosting population by up to two orders of magnitude, since $f_\mathrm{occ}(M_\mathrm{gal,\star})\to0$ is below the $M_0$ scale of Eq.~\eqref{eq:focc}.
The occupation-fraction choice is therefore consequential for the sources generated under the \citet{Hannah:2024dmr} nucleation prescription.
The sources generated under the uniform-occupation assumption should be read as upper limits on the true population.

Once a galaxy is nucleated and hosts an \ac{MBH}, we assign a \ac{MBH} mass using one of two alternatives.

The first method uses the $M_\bullet$--$\sigma$ relation,
\begin{equation}
\label{eq:MBH-sigma}
\log_{10}\left(\frac{M_\bullet}{M_\odot}\right)= 7.87 + 4.55~\log_{10}\!\left(\frac{\sigma}{160~\mathrm{km~s^{-1}}}\right), 
\end{equation}
which has a scatter of $\epsilon = 0.55~\mathrm{dex}$ and the coefficients are taken from \citet{Greene:2019vlv}. 
The velocity dispersion $\sigma$ of a galaxy is calculated as
\begin{equation}
\label{eq:sigma}
\sigma = \sqrt{\frac{GM_\mathrm{gal,\star}}{2 k_\mathrm{vir} R_\mathrm{eff}}},
\end{equation}
where $k_\mathrm{vir}$ is the virial coefficient, which we fix to $k_\mathrm{vir}=1$, and $R_\mathrm{eff}$ is the effective radius, computed as~\citep{Nedkova:2021wkc}
\begin{equation}
\label{eq:Re-Mgal}
\log_{10}\left(\frac{R_\mathrm{eff}}{\mathrm{kpc}}\right) = 0.82 + 0.24\left[\log_{10}\left(\frac{M_\mathrm{gal,\star}}{M_\odot}\right) - 10.7\right] - \alpha_R\log_{10}(1+z), 
\end{equation}
where the scatter is $\epsilon = 0.20~\mathrm{dex}$.
Following \citet{Leigh:2012rd}, we assume $\sigma_\mathrm{NSC}\approx \sigma$, consistent with the approach adopted in \citet{Broggi:2022udp}.
The galaxy size--mass relation evolves with redshift, with galaxies of a given stellar mass being more compact at higher $z$~\citep{vanderWel:2014wba}.
Since $\sigma$ is computed from $M_\mathrm{gal,\star}$ and $R_\mathrm{eff}$ through Eq.~\eqref{eq:sigma}, we include this evolution through the last term of Eq.~\eqref{eq:Re-Mgal}, such that $R_\mathrm{eff}(M_{\mathrm{gal},\star},z)=R_\mathrm{eff}(M_{\mathrm{gal},\star},0)\,(1+z)^{-\alpha_R}$, where we adopt $\alpha_R=0.75$~\citep{vanderWel:2014wba}.

\begin{figure}
    \centering
    \includegraphics[width=\columnwidth]{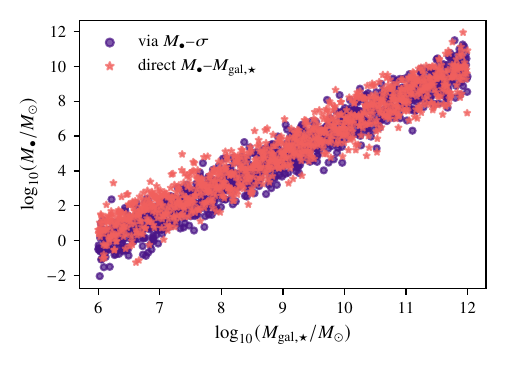}
    \caption{\ac{MBH} mass ($\log_{10}M_\bullet$) as a function of galaxy stellar mass ($M_\mathrm{gal,\star}$) for the sampled galaxy population, assigned by two methods.
    Purple circles show $M_\bullet$ obtained from the $M_\bullet$–$\sigma$ relation, with $\sigma$ derived from the virial relation of Eq.~\eqref{eq:sigma}.
    Pink stars show $M_\bullet$ obtained directly from $M_\mathrm{gal,\star}$ through Eq.~\eqref{eq:MBH-Mgal} for the same galaxies.}
    \label{fig:scaling_relations}
\end{figure}

The second method derives the \ac{MBH} mass $M_\bullet$ directly from the galaxy stellar mass $M_\mathrm{gal,\star}$ via the $M_\mathrm{gal,\star}$--$M_\bullet$ scaling relation of \citet{Greene:2019vlv},
    \begin{equation}
    \label{eq:MBH-Mgal}
        \log_{10}\left(\frac{M_\bullet}{M_\odot}\right) = 7.43 + 1.61\log_{10}\left(\frac{M_\mathrm{gal,\star}}{3\times10^{10}~M_\odot}\right),
    \end{equation}
which has a scatter of $\epsilon = 0.81~\mathrm{dex}$.
We show the resulting scaling of $M_\bullet$ with $M_\mathrm{gal,\star}$ and $\sigma$ in Fig.~\ref{fig:scaling_relations}.
The two methods agree over $\log_{10}(M_\star/M_\odot)\simeq9$--$11$ but diverge at the high- and low-mass ends, affecting the detectable volume of \acp{EMRI} and \acp{TDE} (Sec.~\ref{sec:population_distribution}). 

\subsection{Nuclear stellar cluster evolution and cusp regrowth}
\label{subsec:NSCevolution_cuspregrowth}

Given that we have now evaluated the $M_\bullet$ and $\sigma$, we can set the properties of the \ac{NSC}. 
The \acp{NSC} forming around \acp{MBH} are a consequence of the long-term dynamical evolution of their host galaxies~\citep{Antonini:2015sza,Zhang:2025jmm}. 
We set the influence radius $r_\mathrm{infl}$, the radius up to which the gravitational field from the \ac{MBH} dominates~\citep{2013degn.book.....M}, as
\begin{equation}
\label{eq:r_infl}
    r_\mathrm{infl} = \frac{G M_\bullet}{\sigma^2}.
\end{equation}
In galaxies that have not experienced recent major perturbations, two-body relaxation and mass segregation form a dense stellar cusp in the immediate vicinity of the central \ac{MBH}.
This phase is associated with peak production of \ac{EMRI} and \ac{TDE}.
As the cusp continues to evolve and diffuse outwards, the central stellar density decreases, leading to a decline in the instantaneous rates of both events.

We describe the spatial distribution of the \acp{sBH} and stars in the \ac{NSC} with a Dehnen profile~\citep{Dehnen:1993uh}. 
The number density of a species is 
\begin{equation}
\label{eq:dehnen_eq}
    n_\mathcal{S}(r) = \frac{3-\gamma}{4\pi}~N_i\mathcal{H}(r-r_\mathcal{S})~\frac{r_a}{r^\gamma(r+r_a)^{4-\gamma}},
\end{equation}
Here $N_i$ is the total number of objects of the component, $\mathcal{H}$ is the Heaviside step function (so that the profile is truncated at the inner radius $r_\mathcal{S}$), and $r_a$ is the scale radius.
Following \citet{Broggi:2022udp}, we fix $\gamma=1.5$ for both stars and \acp{sBH} and set  $r_a=4 r_\mathrm{infl}$. 
The inner cut-off radius is different for \acp{sBH} and stars: we use the radius where \acp{sBH} are \emph{captured} ($r_\mathrm{sBH}$) and stars are \emph{disrupted} ($r_\star$).
These can be approximated as 
\begin{equation}
\label{eq:r_k}
\begin{aligned}
r_\mathrm{sBH} &\approx \frac{8GM_\bullet}{c^2}, \\
r_\star &\approx r_{0,\star}\left(\frac{M_\bullet}{m_\star}\right)^{1/3},
\end{aligned}
\end{equation}
where $G$ is the gravitational constant, $c$ is the speed of light, and $m_\star$ and $r_{0,\star}$ are the mass and radius of the disrupted star. 
We adopt $m_\star=1~M_\odot$ and $r_{0,\star}=1\,R_\odot$ through the study.

Given the three-dimensional number density in Eq.~\eqref{eq:dehnen_eq}, we can evaluate the system's profile as
\begin{equation}
\label{density_eq}
    \rho(r) = \sum_\mathcal{S} m_\mathcal{S} n_\mathcal{S}(r) ,
\end{equation}
where $m_\mathcal{S}$ is the mass of the star or \ac{sBH}.
In this study, we restrict ourselves to two components, stars and sBHs. 
The total mass in each component, $M_{\mathrm{tot},\star}=20\,M_\bullet$ and $M_{\mathrm{tot},\mathrm{sBH}}=0.2\,M_\bullet$, is distributed among objects of fixed individual mass, $m_\star=1\,M_\odot$ and $m_\mathrm{sBH}=10~M_\odot$~\citep{Broggi:2022udp}.

The structure of the \ac{NSC} is, however, not static over cosmic time.
Stellar hardening during the evolution of \ac{MBH}--\ac{MBH} mergers erodes a pre-existing cusp~\citep{Babak:2017tow}.
While \ac{MBH}--\ac{MBH} mergers disrupt the \ac{NSC}, here we assume that this is a consequence of major galaxy--galaxy mergers.
Hence, these events can strongly perturb the \ac{NSC}, partially or entirely weakening the pre-existing cusp through violent relaxation and dynamical heating~\citep{Merritt:2006bm}, while at the same time merger-induced nuclear star formation can grow, or regrow a nucleus~\citep{VanWassenhove:2013ota,Pfister:2019rho}.
As a result, the dynamical state of the \ac{NSC} and its ability to produce \acp{EMRI} and \acp{TDE} depend on the time elapsed since the last major galaxy and \ac{MBH} mergers.

For any galaxy observed at $z^{\mathrm{gal}}_{\mathrm{obs}}$, the dynamical state of the cusp is encoded in the \emph{cusp age} for that galaxy.
Major galaxy--galaxy mergers and the ensuing \ac{MBH}--\ac{MBH} mergers play a central role in setting this timescale.
If the \ac{MBH}--\ac{MBH} merger follows the nuclear starburst, which is the case if both merging galaxies host a \ac{MBH} and the \acp{MBH} form a binary, the cusp does not grow instantaneously.
Instead, it regrows over a timescale ranging from several hundred million to a few billion years, depending on the properties of the system~\citep{Amaro-Seoane:2010dzj, 2013degn.book.....M,Babak:2017tow}.
The regrowth of the cusp is primarily driven by collisional relaxation processes within the \ac{NSC}~\citep{Merritt:2005yt}.
This gradually rebuilds a dense central region, redistributing the orbital energy and angular momentum in the cusp.
In gas-rich systems, this process may be accelerated by nuclear star formation, leading to a rapid build-up of a new central stellar component~\citep{Haan:2013iza, Pfister:2019rho}.
We model the cusp age $T_\mathrm{c}$ by accounting for (i) the last \ac{MBH}--\ac{MBH} galaxy merger, which typically erodes or destroys the cusp, and (ii) the subsequent regrowth timescale required to rebuild it.
At the low galaxy masses, where \acp{MBH} may not be ubiquitous~\citep{Volonteri:2007ax, Greene:2019vlv}, cusp destruction caused by \ac{MBH}--\ac{MBH} mergers may not commonly happen, thus making our model conservative.
This cusp age then serves as the key parameter controlling the production of \acp{EMRI} and \acp{TDE} in each galaxy.

We parametrise the galaxy--galaxy major merger rate for mass ratios more equal than $1$:$4$ as~\citep{OLeary:2020edl}
\begin{equation}
\label{eq:galaxy-major-merger}
    \varphi_\mathrm{MM}(z)=\mathcal{R}_0\,(1+z)^\omega\,\exp[\xi(1+z)],
\end{equation}
with the normalisation $\mathcal{R}_0$, power-law slope $\omega$, and exponential-cut-off parameter $\xi$ set by the merger-rate models described in Table~\ref{tab:merger_models}.

\begin{table*}
\centering
\caption{Parameters for the galaxy--galaxy major merger rate $\varphi_\mathrm{MM}(z)$, following the parametrisation of Eq.~\eqref{eq:galaxy-major-merger}, used for validating the merger model, together with the stated validity range of each calibration.}
\label{tab:merger_models}
\begin{tabular}{lcccll}
\hline
Model & Rate normalization $\mathcal{R}_0$~(Gyr$^{-1}$) & Power-law slope $\omega$ & Exponential cut-off $\xi$ & Range of validity ($M_\star$, $z$) \\
\hline
\citet{Mundy:2015qgi} & $1.8\times10^{-2}$ & $1.3$ & $0$ & $z<3.5$, $M_\star/M_\odot >10^{10}$ \\
\citet{OLeary:2020edl} & $10^{-1.785}$ & $1.277$ & $-0.100$ & $z\lesssim6$ , $M_\star/M_\odot \geq10^{9.7}$ \\
\citet{Duan:2025lqz} & $0.013$ & $3.359$ & $-0.138$ & $4.5 \leq z\leq 11.5$, $10^{8} \leq M_\star/M_\odot \leq 10^{10}$ \\
\hline
\end{tabular}
\end{table*}

To connect this merger rate to the evolution of individual \acp{NSC}, we model major mergers as a Poisson process of independent, discrete events.
The expected number of major mergers experienced by a galaxy between two redshifts $z_1$ and $z_2$ is given by
\begin{equation}
\label{eq:cumulative_hazard_function_0}
    \Xi_{12} = \int_{z_1}^{z_2} \varphi_\mathrm{MM}(z)
    \left|\frac{\mathrm{d}t}{\mathrm{d}z}\right| \mathrm{d}z ,
\end{equation}
where the cosmic time $t(z)$ corresponding to redshift $z$ is related via~\citep{Hogg:1999ad}
\begin{equation}
    \frac{\mathrm{d}t}{\mathrm{d}z}=-\frac{1}{(1+z)H_0\sqrt{\Omega_\mathrm{m}(1+z)^3+\Omega_\Lambda}} .
\end{equation}
For a galaxy observed at redshift $z^{\mathrm{gal}}_{\mathrm{obs}}$, we can define the cumulative hazard function,
\begin{equation}
\label{eq:cumulative_hazard_function}
    \Xi(z) = \int_{z^{\mathrm{gal}}_{\mathrm{obs}}}^{z} \varphi_\mathrm{MM}(z')
    \left|\frac{\mathrm{d}t}{\mathrm{d}z'}\right| \mathrm{d}z' ,
\end{equation}
which gives the expected number of major mergers occurring between the observed redshift and a higher redshift $z' > z^{\mathrm{gal}}_{\mathrm{obs}}$ ($t(z') < t(z^{\mathrm{gal}}_{\mathrm{obs}})$).

Within this model, the dynamical state of the \ac{NSC} is directly linked to the time elapsed since the \ac{LMM}.
The absence of mergers over a given time interval is a necessary condition for the cusp to remain intact or to regrow.
The probability that \emph{no} major merger has occurred between redshift $z'$ and observed $z^{\mathrm{gal}}_{\mathrm{obs}}$ is given by the survival function
\begin{equation}
\label{eq:survival_function}
    S(z') = \exp\left[-\Xi(z')\right].
\end{equation}
This survival probability $S(z')$ represents that the \ac{NSC} has not been disrupted by a major merger since redshift $z'$.

The probability density that the \ac{LMM} occurred at redshift $z_\mathrm{LMM}>z^{\mathrm{gal}}_{\mathrm{obs}}$ is therefore obtained by combining two requirements.
First, that a merger occurs at $z_{\mathrm{LMM}}$, and second, that no mergers occur at later times (i.e. at redshifts between $z^{\mathrm{gal}}_{\mathrm{obs}}$ and $z_{\mathrm{LMM}}$).
The first condition is specified by the instantaneous merger rate, while the second is enforced by the survival function.
The resulting probability density function is
\begin{equation}
\label{eq:p_LMM}
    p_{\mathrm{LMM}}(z_{\mathrm{LMM}}\mid z^{\mathrm{gal}}_{\mathrm{obs}})
    = \varphi_{\mathrm{MM}}(z_{\mathrm{LMM}})
    \left|\frac{\mathrm{d}t}{\mathrm{d}z_\mathrm{LMM}}\right|
    \exp\!\left[-\Xi(z_{\mathrm{LMM}})\right],
\end{equation}
normalised over the interval $z_{\mathrm{LMM}}\in[z,\,z_{\max}]$.

The newly formed cusp regrowth time is of order one local relaxation time at the \ac{MBH} influence radius~\citep{2013degn.book.....M, Babak:2017tow}.
Given the \ac{NSC} profile, we can evaluate the relaxation time $t_\mathrm{rlx}$~\citep{Elson:1986sm, 2013degn.book.....M} as 
\begin{equation}
\label{relax_time_full_eq}
t_{\mathrm{rlx}}(r)\simeq\frac{0.34\sigma^3}{G^2\bar m(r) \rho(r)\ln\Lambda_\mathrm{C}},
\end{equation}
where $\ln\Lambda_\mathrm{C}$ is the Coulomb logarithm  and $\bar{m}$ is the first moment of the mass density.
The Coulomb logarithm accounts for the cumulative effect of many small-angle gravitational encounters~\citep{Elson:1986sm, Chandrasekhar:1943ys}, and we fix it to $15$~\citep{Merritt:2009mr}. 
We define $\bar{m}$ as~\citep{Elson:1986sm}
\begin{equation}
\label{eq:mbar}
\bar{m}(r) = \frac{ m_{\star}^2 \, n_{\star}(r) + m_{\mathrm{sBH}}^2 \, n_{\mathrm{sBH}}(r)}
{m_{\star} \, n_{\star}(r) + m_{\mathrm{sBH}} \, n_{\mathrm{sBH}}(r)}.
\end{equation}
To calculate the relaxation timescale of the \ac{NSC}, we set $r=r_{\mathrm{infl}}$ for Eq.~\eqref{relax_time_full_eq}.

Since cusp regrowth requires a finite period of relaxation following the last major merger, we define a cusp turn-on time as
\begin{equation}
\label{eq:t_on}
    t_{\mathrm{on}}=t_\mathrm{LMM}+\kappa t_\mathrm{rlx}(r_{\mathrm{infl}}).
\end{equation}
Here $\kappa$ parametrises the fraction of the relaxation time required for the stellar cusp to re-establish itself in a dynamically relaxed configuration across the central region relevant for \ac{EMRI} and \ac{TDE} production.
Following \citet{Babak:2017tow}, we adopt $\kappa = 0.25$.

Finally for a galaxy observed at $z^{\mathrm{gal}}_{\mathrm{obs}}$, with cosmic time $t(z^{\mathrm{gal}}_{\mathrm{obs}})$, the cusp age is
\begin{equation}
\label{eq:cusp_age}
    T_\mathrm{c}(z^{\mathrm{gal}}_{\mathrm{obs}}) = \max\left\{0,\ t(z^{\mathrm{gal}}_{\mathrm{obs}})-t_\mathrm{ on}\right\}.
\end{equation}

\begin{figure}
    \centering
    \includegraphics[width=\columnwidth]{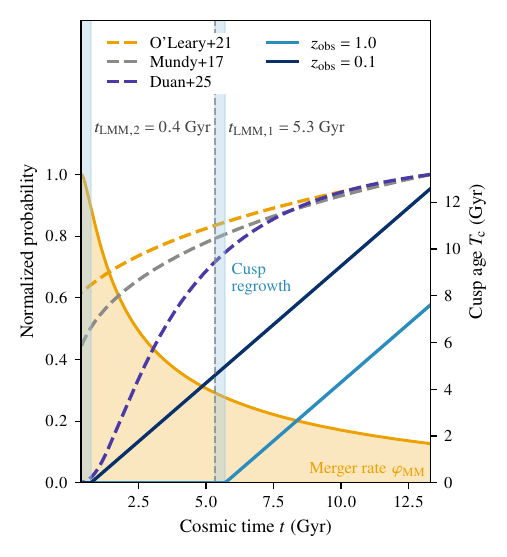}
    \caption{Evolution of \ac{NSC} properties after galaxy--galaxy major mergers for two redshifts.
    The orange curve shows the normalised major merger rate $\varphi_\mathrm{MM}(z)$ of our fiducial model, \citet{OLeary:2020edl}. 
    The dashed curves show the survival function $S(z)$ for the three merger models (Table~\ref{tab:merger_models})
    The vertical grey dashed lines mark the sampled last-major-merger times $t_\mathrm{LMM}$ of the two galaxies at different redshifts, each followed by a shaded cusp regrowth phase of duration $\kappa t_\mathrm{rlx}$ during which the cusp age is zero.
    Thereafter, the cusp age $T_\mathrm{c}=t-t_\mathrm{on}$ grows linearly.
    The $z_\mathrm{obs}=1.0$ galaxy has a sampled merger at $t_{\mathrm{LMM},1}=5.3~\mathrm{Gyr}$, whereas the $z_\mathrm{obs}=0.1$ galaxy has $t_{\mathrm{LMM},2}=0.4~\mathrm{Gyr}$, close to the earliest epoch simulated ($z_\mathrm{max}=12$), and therefore hosts a maximally old cusp at the time of observation.}
    \label{fig:nsc_cusp_evolution}
\end{figure}

In Fig.~\ref{fig:nsc_cusp_evolution}, we illustrate how the redshift evolution of the galaxy major merger rate gives rise to a broad distribution of \ac{NSC} dynamical ages.
Since mergers are more frequent at early cosmic times, systems that have not undergone recent mergers can evolve uninterrupted for longer periods, allowing their stellar cusps to grow and relax by the observation epoch.
Conversely, galaxies with more recent mergers undergo cusp disruption followed by a regrowth phase, which delays the onset of sustained dynamical relaxation.

\subsection{Simulating EMRIs and TDEs}
\label{subsec:simulating_EMRIs_TDEs}

\begin{figure}
    \centering
    \includegraphics[width=\columnwidth]{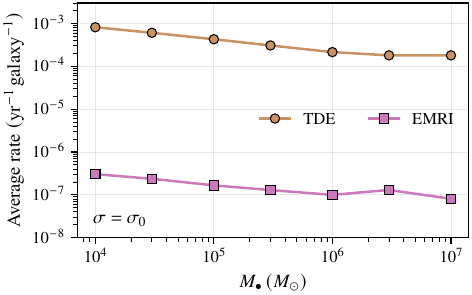}
    \caption{Average per-galaxy \ac{EMRI} (squares) and \ac{TDE} (circles) rates as a function of \ac{MBH} mass $M_\bullet$ at $\sigma=\sigma_0$, derived from the simulations of \citet{Broggi:2022udp}.}
    \label{fig:average_rates}
\end{figure}

To model the production of \acp{EMRI} and \acp{TDE} in galactic nuclei, we use the simulation outputs of \citet{Broggi:2022udp} to compute the average rate over their simulations.
This choice allows us to build a general strategy that is less dependent on specific simulations.
We consider only \acp{EMRI} formed through two-body relaxation in \acp{NSC} (dry \acp{EMRI}) and do not model \acp{EMRI} formed in \ac{AGN} discs~\citep[\emph{wet} \acp{EMRI};][]{Lyu:2024gnk, Pan:2021oob}.
For each combination of \ac{MBH} mass and nuclear density, \citet{Broggi:2022udp} report the total number of \acp{EMRI} ($N_\mathrm{EMRI}$) and \acp{TDE} ($N_\mathrm{TDE}$) produced over the lifetime of the simulated cusp, together with a characteristic \ac{EMRI} timescale $t_\mathrm{E}$.
We use these to construct a two-dimensional grid in $M_\bullet$ and $\sigma/\sigma_0$, spanning $M_\bullet \in [10^4,\,10^7]~M_\odot$ and four density levels corresponding to $\sigma/\sigma_0 = (\rho/\rho_0)^{1/6} \in \{0.464,\,0.681,\,1.0,\,1.468\}$, where $\sigma_0(M_\bullet)$ follows the scaling of~\citet[Eq.~15]{Broggi:2022udp}.

For each grid point, we define an effective duration over which the cusp actively produces events,
\begin{equation}
\label{eq:t_f}
    t_\mathrm{f} = \min\left\{10\,t_\mathrm{E},\ 10~\mathrm{Gyr}\right\},
\end{equation}
and the corresponding constant per-galaxy rates
\begin{equation}
\label{eq:constant_rate}
    \Gamma_{\mathrm{EMRI}} = \frac{N_\mathrm{E}}{t_\mathrm{f}}, \qquad
    \Gamma_{\mathrm{TDE}} = \frac{N_\mathrm{T}}{t_\mathrm{f}}.
\end{equation}
We interpolate $\log_{10}\Gamma_\mathcal{S}$ bilinearly on the $(\log_{10}M_\bullet,\ \log_{10}(\sigma/\sigma_0))$ grid for each galaxy in our catalogue, clipping both axes to the grid boundaries to avoid extrapolation.
In Fig.~\ref{fig:average_rates} we show the resulting rates as a function of $M_\bullet$ for $\sigma=\sigma_0$.
Both rates decrease with $M_\bullet$, with \acp{TDE} exceeding \acp{EMRI} by roughly three orders of magnitude.

Under this model, a galaxy produces events at a rate $\Gamma_\mathcal{S}$ only while its cusp remains dynamically active, which is when the cusp age $T_\mathrm{c}$ is shorter than the local relaxation time $t_{\mathrm{rlx}}(r_\mathrm{infl})$.
The expected number of events of type $k$ observed over an observation window $T_{\mathrm{obs}}$ is therefore
\begin{equation}
\label{eq:observed_events_in_time}
    N_\mathcal{S}(z) = \Gamma_\mathcal{S}\,\frac{T_{\mathrm{obs}}}{1+z}\mathcal{H}\left(t_{\mathrm{rlx}} - T_\mathrm{c}\right),
\end{equation}
where the factor $(1+z)$ accounts for cosmological time dilation.
The generated number of events for each galaxy is drawn from a Poisson distribution, $N_\mathcal{S}^{\mathrm{generated}} \sim \mathrm{Poisson}(N_\mathcal{S})$.

The observation duration $T_{\mathrm{obs}}$ in Eq.~\eqref{eq:observed_events_in_time} depends on the detector under consideration.
For \acp{EMRI}, we adopt an optimistic \ac{LISA} mission lifetime of $T_{\mathrm{obs}}^{\mathrm{EMRI}} = 6~\mathrm{yr}$~\citep{LISA:2024hlh}.
For \acp{TDE}, we consider the combined baseline of \ac{ZTF} and \ac{LSST}.
\ac{ZTF} survey has been operating since 1 March 2018 (MJD~58178) and we take its observation window to extend through December 2026 (MJD~61396), corresponding to $T_{\mathrm{obs}}^{\mathrm{ZTF}} = 8.75~\mathrm{yr}$~\citep{Bellm:2018cps}, while \ac{LSST} is assumed to begin science operations in June 2025 (MJD~60827) and run through June 2035 (MJD~64423), corresponding to $T_{\mathrm{obs}}^{\mathrm{LSST}} = 9.84~\mathrm{yr}$~\citep{LSST:2008ijt}.
Combined, the two surveys span a total \ac{TDE} observation baseline of $T_{\mathrm{obs}}^{\mathrm{TDE}} = 17.25~\mathrm{yr}$, from the start of \ac{ZTF} to the end of \ac{LSST}.

\subsection{Properties of the generated sources}
\label{subsec:EMRI_TDE_properties}

The astrophysical framework described in Sec.~\ref{sec:astrophysical_setup} assigns each event a host galaxy and, consequently, a set of global properties, including the \ac{MBH} mass ($M_\bullet$), sky position and redshift ($z_{\mathrm{gal}}$).
These quantities are therefore determined self-consistently by the galaxy population and merger-driven evolution.
To complete the description, we specify the remaining intrinsic parameters entering the observable signals.
The generic forms of the distributions used in this work are presented in Table~\ref{tab:generic_form}.

We model the spin parameter of the \acp{MBH} $a_\bullet$ using a bimodal prescription motivated by different growth channels~\citep{Fanidakis:2010dd}.
We adopt a double Beta distribution in spin magnitude, with characteristic peaks at $a_\bullet \simeq 0.3$ for $M_\bullet \lesssim 10^{6}~M_\odot$ and $a_\bullet \simeq 0.7$ for $M_\bullet \gtrsim 10^{6}~M_\odot$.
This choice reflects the expectation that lower-mass \acp{MBH} may preferentially grow through stochastic, low-angular-momentum accretion episodes, while higher-mass systems are more likely to experience prolonged, coherent accretion or merger-driven spin-up~\citep{King:2008au, Berti:2008af}, while at the highest masses spin is expected to decrease to $\sim 0.7$ because their host galaxies are at the same time gas poor and experiencing many mergers, MBH mergers are thus suggested to dominate recent growth of these MBHs \citep{Dubois:2013rha}.  

For the secondary component, we assume fixed masses, with \acp{sBH} having $m_{\mathrm{sBH}} = 10~M_\odot$ and stars having $m_\star = 1~M_\odot$, consistent with the choices adopted in Sec.~\ref{subsec:galactic_properties}.

\begin{figure}
    \centering
    \includegraphics[width=\columnwidth]{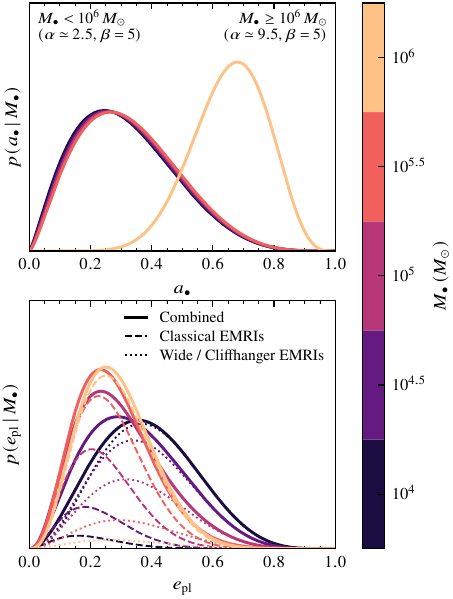}
    \caption{Functional forms of the \ac{MBH} spin ($a_\bullet$; top) and plunge-eccentricity ($e_\mathrm{pl}$; bottom) as a function of \ac{MBH} mass used in catalogue source generation (Table~\ref{tab:simulated_catalogue_scenario_summary}).}
    \label{fig:emri_ecc_spin_vs_MBH}
\end{figure}

\acp{EMRI} formed in \acp{NSC} are expected to retain non-negligible eccentricities when they enter the \ac{GW}-driven regime. 
We denote by $e_\mathrm{pl}$ the orbital eccentricity at the onset of the final plunge.
Recent numerical studies have shown that the distribution of $e_{\mathrm{pl}}$ is broad and exhibits a significant high-eccentricity tail, with a peak around $e_{\mathrm{pl}} \sim 0.2$ and a non-negligible fraction of systems (up to $\sim20\%$ for $M_\bullet\sim10^{5}$--$10^{6}~M_\odot$) reaching $e_{\mathrm{pl}}\gtrsim 0.5$ ~\citep{Mancieri:2025cmx}.
These features arise naturally from the interplay between two-body relaxation and gravitational-wave emission.
This gives rise to two distinct dynamical channels.
The \emph{classical} channel produces systems with moderate eccentricities.
It comprises \acp{EMRI} formed from orbits with semi-major axes below a critical scale $a_\mathrm{c}$, beyond which captures were traditionally expected to end almost exclusively as direct plunges.
The \emph{wide} channel populates the high-eccentricity tail.
It is made up of \emph{cliffhanger} \acp{EMRI}, which form at semi-major axes well beyond $a_\mathrm{c}$, appear predominantly for $M_\bullet\lesssim3\times10^{5}~M_\odot$, and may account for up to $\sim55\%$ of all \acp{EMRI} in the lowest-mass \acp{MBH}~\citep{Mancieri:2024sfy}.
To incorporate these results into our population synthesis model, we parametrise the distribution $p(e_{\mathrm{pl}})$ as a bimodal Beta distribution whose two modes correspond to the classical and cliffhanger channels, respectively (Fig.~\ref{fig:emri_ecc_spin_vs_MBH}).
The shape parameters and the mixing fraction are functions of \ac{MBH} mass, calibrated against the simulated plunge-eccentricity histograms in \citet{Mancieri:2025cmx}.

Our baseline assumption for the inclination angle $\iota_0$ of \acp{EMRI} is that it is uniformly distributed in $\cos \iota_0 \in [-1, 1]$, where positive (negative) values correspond to prograde (retrograde) orbits.
This is a common choice~\citep{Amaro-Seoane:2007osp, Gair:2010yu} for isotropic populations. 

Since \ac{LISA} observes a continuous population of \acp{EMRI} at various stages of inspiral, not every source is guaranteed to plunge within the observation window.
To reflect this realism, we assign a plunge time $T_{\mathrm{plunge}}$ to each source, uniformly drawn between $1~\mathrm{yr}$ and $6~\mathrm{yr}$~\citep{Singh:2025pzh}.
This allows a source to plunge at any point during, or after, the \ac{LISA} observation window, mimicking the realistic condition that a source's inspiral is not synchronised with the beginning of the mission and that only a subset of the population is expected to complete its inspiral while \ac{LISA} is observing~\citep{Oliver:2026krg}.
Following previous work~\citep{Chapman-Bird:2022tvu, Singh:2025pzh}, we draw the initial azimuthal, polar and radial phases, $\Phi_{\phi,0}$, $\Phi_{\theta,0}$ and $\Phi_{r,0}$, independently and uniformly in $[0, 2\pi)$, consistent with the absence of any preferred orbital phase at the start of the observation window.

For \acp{TDE}, we characterise each disruption event using three key parameters: the orbital penetration factor $\beta_{\mathrm{orb}}$, the disc viscosity parameter $\alpha_{\mathrm{visc}}$, and the feedback efficiency $\eta$.
The penetration factor $\beta_{\mathrm{orb}} \equiv r_\star/r_{\mathrm{p}}$ quantifies the depth of the stellar encounter relative to the tidal radius $r_\star$, defined in Eq.~\eqref{eq:r_k}, where $r_\mathrm{p}$ is the orbit's pericentre distance.
A value $\beta_{\mathrm{orb}} = 1$ corresponds to a grazing encounter at the classical tidal disruption threshold~\citep{Rees:1988bf}, while larger values indicate progressively deeper plunges, with the star penetrating well within the tidal radius and experiencing stronger tidal forces.
We restrict $\beta_{\mathrm{orb}} \in [0.6,4]$, encompassing the full range from marginal full disruptions to deep encounters, which may result in partial or complete stripping of the stellar envelope and significantly affect the resulting debris fallback rate~\citep{Guillochon:2012uc}.

The viscosity parameter $\alpha_{\mathrm{visc}}$ enters through the standard Shakura--Sunyaev framework~\citep{Shakura:1972te} for the $\alpha$-disc prescription and governs the rate at which the circularised debris disc transports angular momentum outwards, regulating the accretion rate onto the \ac{MBH}.
The feedback efficiency $\eta$ quantifies the fraction of the accreted rest-mass energy that is radiated or deposited back into the surrounding environment, thereby controlling the flare's bolometric luminosity~\citep{Metzger:2022lob}.
Both $\alpha_{\mathrm{visc}}$ and $\eta$ are subject to considerable theoretical uncertainty; we therefore adopt broad, uninformative priors $\alpha_{\mathrm{visc}} \in [10^{-2},\,10^{-1}]$ and $\eta \in [10^{-3},\,10^{-1}]$, consistent with the ranges explored in \citet{Sarin:2023uqa}.
These ranges span standard values for $\alpha$-disc viscosity and encompass sub-canonical to near-canonical radiative efficiencies, relevant to the \ac{TDE} disc phase~\citep{King:2007cu}.
Since $\alpha_{\mathrm{visc}}$ and $\eta$ are uncertain at the order-of-magnitude level, both are sampled log-uniformly over their respective ranges.
Independent of these three parameters, the disruption epoch of each event is uniformly sampled over the combined survey baseline.
This mimics realistic conditions in which part of the light curve may not be observed because its total duration falls outside the observing window.

Disruption epochs are drawn uniformly in MJD across the combined \ac{ZTF}+\ac{LSST} baseline. 
We restrict redshifts to $z\leq0.5$ for \ac{ZTF} and $z\leq4$ for \ac{LSST}, beyond which no detections are expected~\citep{Bellm:2018cps, Karmen:2026ixp}.

\begin{table*}
    \centering
    \footnotesize
    \caption{Parametric forms used to model the underlying population distributions in this work.
    Each parameter ${\theta}_j \in \boldsymbol{\theta}$ is sampled from a distribution $p_{\mathrm{pop}}({\theta}_j|\boldsymbol{\Lambda})$, characterized by hyperparameters $\boldsymbol{\Lambda}$.}
    \label{tab:generic_form}
    \begin{tabular}{lcc}
        \hline
        Distribution & Functional form $p_{\mathrm{pop}}({\theta}_j|\boldsymbol{\Lambda})$ & Hyperparameters $\boldsymbol{\Lambda}$ \\
        \hline
        Beta &
        ${\theta}_j^{\alpha-1}(1-{\theta}_j)^{\beta-1}$ &
        $\alpha,\beta$ \\

        Truncated-normal &
        $\exp[-({\theta}_j-\mu)^2/(2\sigma^2)]$ &
        $\mu,\sigma$ \\

        Power-law &
        ${\theta}_j^\lambda$ &
        $\lambda$ \\

        Schechter &
        $x_c^{-1}\left({\theta}_j/x_c\right)^{\zeta}
        \exp\!\left(-{\theta}_j\zeta/x_c\right)$ &
        $\zeta,x_c$ \\

        Skew-normal &
        $\exp(-{\theta}_j^2/2)\,
        [1+\erf(\gamma{\theta}_j/\sqrt{2})]$ &
        $\gamma$ \\

        Bimodal Beta &
        $f{\theta}_j^{\alpha_1-1}(1-{\theta}_j)^{\beta_1-1}
        +(1-f){\theta}_j^{\alpha_2-1}(1-{\theta}_j)^{\beta_2-1}$ &
        $f,\alpha_1,\beta_1,\alpha_2,\beta_2$ \\

        Bimodal Gaussian &
        $f\exp[-({\theta}_j-\mu_1)^2/(2\sigma_1^2)]
        +(1-f)\exp[-({\theta}_j-\mu_2)^2/(2\sigma_2^2)]$ &
        $f,\mu_1,\sigma_1,\mu_2,\sigma_2$ \\

        \hline
    \end{tabular}
\end{table*}

\subsection{Simulated catalogues}
\label{sec:simulated_catalogue}

Combining all cases for the scaling relation, nucleation fraction, and occupation probability yields 8 population scenarios.
Only galaxies that are simultaneously nucleated \emph{and} occupied are carried forward into the \ac{EMRI}/\ac{TDE} rate calculation.

For each of the 8 scenarios defined in Sec.~\ref {subsec:galactic_properties}, we generate $N=100$ independent Monte Carlo realisations and compute the number of \ac{EMRI} and \ac{TDE} detected sources.
The scenarios are labelled by a three-part code, as summarised in Table~\ref{tab:simulated_catalogue_scenario_summary}: the \ac{MBH}-mass relation, $\sigma$ for $M_\bullet$--$\sigma$, $\star$ for $M_\bullet$--$M_\star$; the nucleation model \texttt{H} for \citet{Hannah:2024dmr}, \texttt{N} for \citet{Neumayer:2020gno}, and the occupation model, \texttt{U} for uniform, \texttt{V} for varying. 
We select a single median realisation and use it for all subsequent analysis. 
For each scenario, we summarise the per-source detected counts across the $N=\simNumRealisations$ realisations within their $90\%$ intervals (Table~\ref{tab:simulated_catalogue_scenario_summary}) and present the universe $\mathcal{U}^\star$ that we used as the input for the downstream population analysis.
The detected number of sources in $\mathcal{U}^\star$ is then representative of the underlying catalogue.


\begin{table*}
\centering
\footnotesize
\setlength{\tabcolsep}{3pt}
\renewcommand{\arraystretch}{1.45}
\caption{Population scenarios and detection summary.
Each scenario is the cross product of the \ac{MBH}-mass relation, nucleation fraction, and the occupation fraction (Sec.~\ref{sec:simulated_catalogue}).
Detected counts are medians over $N=\simNumRealisations$ independent realisations for each channel.
The $\mathcal{U}^\star$ columns give the counts for the single realisation used in all downstream cross-scenario comparisons. 
The final two columns give the median \ac{EMRI}-to-\ac{TDE} detection ratio.
Uncertainties are the equal-tailed $90\%$ interval across realisations. 
The \ac{EMRI} results are presented for a $6$~yr \ac{LISA} mission; the \ac{TDE} count for \ac{ZTF} assumes an $8.75$~yr survey and $9.84$~yr for \ac{LSST}.}
\label{tab:simulated_catalogue_scenario_summary}
\begin{tabular*}{\textwidth}{@{\extracolsep{\fill}}c c c c c c c c c c c c}
\hline
 & \multicolumn{3}{c}{Scenario definition} & \multicolumn{6}{c}{Detected counts} & \multicolumn{2}{c}{EMRI/TDE ratio} \\
 \cline{2-4} \cline{5-10} \cline{11-12}
Model & $M_\bullet$ relation & Nucleation & $f_\mathrm{occ}$ & EMRI & $\mathcal{U}^\star$ & TDE (ZTF) & $\mathcal{U}^\star$ & TDE (LSST) & $\mathcal{U}^\star$ & ZTF ($\times10^{-2}$) & LSST ($\times10^{-4}$) \\
\hline
$\star\texttt{HU}$ & $M_\bullet$--$M_\star$ & Hannah   & Uniform & \tabCountEmriStarHU & \tabUstarEmriStarHU & \tabCountZtfStarHU & \tabUstarZtfStarHU & \tabCountLsstStarHU & \tabUstarLsstStarHU & \tabRatioZtfStarHU & \tabRatioLsstStarHU \\
$\sigma\texttt{HU}$ & $M_\bullet$--$\sigma$ & Hannah   & Uniform & \tabCountEmriSigHU & \tabUstarEmriSigHU & \tabCountZtfSigHU & \tabUstarZtfSigHU & \tabCountLsstSigHU & \tabUstarLsstSigHU & \tabRatioZtfSigHU & \tabRatioLsstSigHU \\
$\star\texttt{NU}$ & $M_\bullet$--$M_\star$ & Neumayer & Uniform & \tabCountEmriStarNU & \tabUstarEmriStarNU & \tabCountZtfStarNU & \tabUstarZtfStarNU & \tabCountLsstStarNU & \tabUstarLsstStarNU & \tabRatioZtfStarNU & \tabRatioLsstStarNU \\
$\sigma\texttt{NU}$ & $M_\bullet$--$\sigma$ & Neumayer & Uniform & \tabCountEmriSigNU & \tabUstarEmriSigNU & \tabCountZtfSigNU & \tabUstarZtfSigNU & \tabCountLsstSigNU & \tabUstarLsstSigNU & \tabRatioZtfSigNU & \tabRatioLsstSigNU \\
$\star\texttt{HV}$ & $M_\bullet$--$M_\star$ & Hannah   & Varying & \tabCountEmriStarHV & \tabUstarEmriStarHV & \tabCountZtfStarHV & \tabUstarZtfStarHV & \tabCountLsstStarHV & \tabUstarLsstStarHV & \tabRatioZtfStarHV & \tabRatioLsstStarHV \\
$\sigma\texttt{HV}$ & $M_\bullet$--$\sigma$ & Hannah   & Varying & \tabCountEmriSigHV & \tabUstarEmriSigHV & \tabCountZtfSigHV & \tabUstarZtfSigHV & \tabCountLsstSigHV & \tabUstarLsstSigHV & \tabRatioZtfSigHV & \tabRatioLsstSigHV \\
$\star\texttt{NV}$ & $M_\bullet$--$M_\star$ & Neumayer & Varying & \tabCountEmriStarNV & \tabUstarEmriStarNV & \tabCountZtfStarNV & \tabUstarZtfStarNV & \tabCountLsstStarNV & \tabUstarLsstStarNV & \tabRatioZtfStarNV & \tabRatioLsstStarNV \\
$\sigma\texttt{NV}$ & $M_\bullet$--$\sigma$ & Neumayer & Varying & \tabCountEmriSigNV & \tabUstarEmriSigNV & \tabCountZtfSigNV & \tabUstarZtfSigNV & \tabCountLsstSigNV & \tabUstarLsstSigNV & \tabRatioZtfSigNV & \tabRatioLsstSigNV \\
\hline
\end{tabular*}
\end{table*}

\begin{figure}
    \centering
    \includegraphics[width=\linewidth]{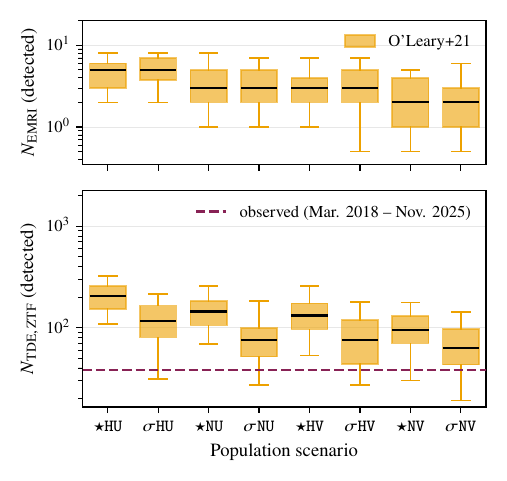}
    \caption{Impact of the merger-rate model choice on the detected \ac{EMRI} (\emph{top}) and \ac{TDE}-\ac{ZTF} (\emph{bottom}) counts, for each of the eight population scenarios (Table~\ref{tab:simulated_catalogue_scenario_summary}), for $\simNumRealisations{}$ realisations per scenario for the \citet{OLeary:2020edl} merger model. 
    Other merger models (Table~\ref{tab:merger_models}) are shown in Fig.~\ref{fig:merger_model_impact_mundy_duan}.
    Boxes show the interquartile range with the median indicated by the horizontal black line; whiskers extend to the 5th and 95th percentiles.
    Detections are shown for a $6~\mathrm{yr}$ \ac{LISA} mission for \acp{EMRI} and for $8.75~\mathrm{yr}$ \ac{ZTF}-\acp{TDE} observations.
    The dashed line shows  confirmed \ac{ZTF}-\acp{TDE} detections~\citep{vanVelzen:2020cwu, Hammerstein:2022wia, Yao:2023rbr}.
}
    \label{fig:merger_model_impact_oleary}
\end{figure}

\begin{figure}
    \centering
    \includegraphics[width=\columnwidth]{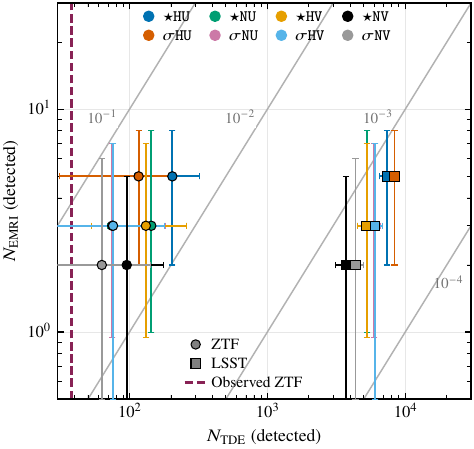}
    \caption{Number of detected \acp{EMRI} against the number of detected \acp{TDE} for the eight population scenarios (Table~\ref{tab:simulated_catalogue_scenario_summary}) under the fiducial \citet{OLeary:2020edl} merger-rate model.
    Each marker shows the median over the $\simNumRealisations{}$ realisations of a scenario, with error bars spanning the 5th to 95th percentiles in both counts; circles use \ac{ZTF} \acp{TDE} and squares use \ac{LSST} \acp{TDE}, and colours identify the scenario.
    Grey lines are lines of constant $N_\mathrm{EMRI}/N_\mathrm{TDE}$.
    The dashed vertical line marks the number of \acp{TDE} observed by \ac{ZTF} \citep{Hammerstein:2022wia}.
    The scenario medians correspond to $\simEmriPerTdeZtfMin$--$\simEmriPerTdeZtfMax$ \acp{EMRI} per \ac{ZTF} \ac{TDE} and $(\simEmriPerTdeLsstMinTimesTenThousand$--$\simEmriPerTdeLsstMaxTimesTenThousand)\times10^{-4}$ per \ac{LSST} \ac{TDE}.}
    \label{fig:emri_to_tde}
\end{figure}

Across the $\simNumRealisations$ realisations, the scenarios predict varying detected counts for \acp{EMRI}, \ac{ZTF}-\acp{TDE} and \ac{LSST}-\acp{TDE} (Table~\ref{tab:simulated_catalogue_scenario_summary}).
The scenarios differ in the \ac{MBH}-mass relation, nucleation fraction and occupation fraction, so the spread in these counts measures how sensitive the event rates are to each of these choices. 
The variation in detection rates across the scenarios is shown in Fig.~\ref{fig:merger_model_impact_oleary}. 
Despite the variation in assumptions leading to variation in total detections, the ratio of \ac{EMRI} to \ac{TDE} detections remains comparable between scenarios, as illustrated in Fig.~\ref{fig:emri_to_tde}.

The \ac{EMRI} and \ac{LSST} counts are forecasts, but the \ac{ZTF} counts can already be compared with an observed sample.
We therefore use the observed \ac{ZTF}-\ac{TDE} count as an independent test of the scenarios before performing any hierarchical inference.
The results are shown in Fig.~\ref{fig:merger_model_impact_oleary} and Fig.~\ref{fig:emri_to_tde}.
The \simObservedZtfTde{} observed \ac{ZTF}-\acp{TDE} between March 2018 and November 2025 already discriminate between the catalogue models.
Every scenario predicts a median \ac{ZTF} detection count above the observed value.
Scenarios with a direct $M_\bullet$--$M_{\rm gal,\star}$ relation and uniform occupation ($\star\texttt{HU}$, $\star\texttt{NU}$) are strongly disfavoured.
The best agreement is found for the $M_\bullet$--$\sigma$ scenarios with either a lower nucleation fraction~\citep{Neumayer:2020gno} or a varying occupation fraction~\citep{Zou:2025mdy}, observed for $\sigma\texttt{NU}$, $\sigma\texttt{HV}$ and $\sigma\texttt{NV}$, for which $16$--$18\%$ of universes lie at or below the observed count.
Combining the \citet{Hannah:2024dmr} nucleation fraction with uniform occupation overpredicts the rate even for the $M_\bullet$--$\sigma$ relation ($\sigma\texttt{HU}$), and requires a varying occupation fraction to match the observations.

\section{Joint Hierarchical Inference}
\label{sec:Joint_Hierarchical_Inference}

The goal of the joint hierarchical inference is to estimate the underlying astrophysical distribution from independent catalogues of $N_{\mathrm{det}}^{\mathrm{EMRI}}$ \ac{EMRI} detections and $N_{\mathrm{det}}^{\mathrm{TDE}}$ \ac{TDE} detections.
This analysis requires several ingredients. 
We must have: a signal model that for each detections maps between observations and source properties such as the \ac{MBH} mass (Sec.~\ref{sec:signal-models}); an estimate for how well the source properties can be measured from each signal (Sec.~\ref{sec:single-pe}); a model for the underlying astrophysical distribution of sources (Sec.~\ref{sec:population_distribution}); a model for observational selection effects (Sec.~\ref{sec:selection-function}) to convert between the observed catalogue of sources and the underlying distribution, and an inference analysis to combine all the observations together and provide constraints on the parameters describing astrophysical population (Sec.~\ref{sec:inference-setup}).

We model the underlying source population with a set of \emph{hyperparameters} $\boldsymbol{\Lambda}$ that describe the population distribution, and we infer their values from the observed catalogue accounting for the parameter uncertainties of individual events.
Using Bayes' theorem, we can express the \emph{hyperposterior} (population-level posterior) as 
\begin{equation}
\label{eq:Hierarchical_Inference_eq}
p(\boldsymbol{\Lambda}|\boldsymbol{d}) = \frac{\pi(\boldsymbol{\Lambda})\mathcal{L}(\boldsymbol{d}|\boldsymbol{\Lambda})}{\mathcal{Z}(\boldsymbol{d})},
\end{equation}
where in the numerator, the first term $\pi(\boldsymbol{\Lambda})$ is the \emph{hyperprior} for the population parameters, and the second term is the \emph{hyperlikelihood} $\mathcal{L}$ for our set of observed data $\boldsymbol{d}$, and the denominator is the \emph{hyperevidence} $\mathcal{Z}$.
Our data $d$ here is the uncorrelated data from the two sources $\boldsymbol{d} = \{\boldsymbol{d}_{\mathrm{EMRI}},\, \boldsymbol{d}_{\mathrm{TDE}}\}$.

Each observed detection $i$ is described by source parameters $\boldsymbol{\theta}_i$ (Table~\ref{tab:models}).
Under the assumption that detections are independent given a population model $p_{\mathrm{pop}}(\boldsymbol{\theta}|\boldsymbol{\Lambda})$, the hyperlikelihood can be expressed by marginalising over $\boldsymbol{\theta}_i$~\citep{Mandel:2018mve, Vitale:2020aaz} as
\begin{equation}
\label{eq:hyperlikelihood_k}
\begin{split}
    \mathcal{L}(\boldsymbol{d} \mid \boldsymbol{\Lambda})
    = & {} \prod_{\mathcal{S}} e^{-\mathcal{R}_\mathcal{S} \alpha_\mathcal{S}(\boldsymbol{\Lambda})}
      \left[\mathcal{R}_\mathcal{S} \alpha_\mathcal{S}(\boldsymbol{\Lambda})\right]^{N_{\mathrm{det}}^\mathcal{S}} \\
     & {} \times\prod_{i=1}^{N_{\mathrm{det}}^\mathcal{S}}
      \frac{\int \mathrm{d}\boldsymbol{\theta}_i\,
            p(d_i \mid \boldsymbol{\theta}_i)\,
            p_{\mathrm{pop}}(\boldsymbol{\theta}_i \mid \boldsymbol{\Lambda})}
           {\alpha_\mathcal{S}(\boldsymbol{\Lambda})}.
\end{split}
\end{equation}
Here, the first product is across $\mathcal{S}$ source types: \ac{EMRI} and \ac{TDE}.
The factorisation of the hyperlikelihood into separate \ac{EMRI} and \ac{TDE} contributions is exact under the assumption of independent noise, and holds regardless of whether the two source populations share a common source population.
The second product accounts for the marginalised likelihood of each detection ($N_{\mathrm{det}}^\mathcal{S}$) from the chosen population model $p_\mathrm{pop}(\boldsymbol{\theta}|\boldsymbol{\Lambda})$: the likelihood $p(d_i \mid \boldsymbol{\theta}_i)$ encodes how well source properties can be measured from each detection, while the population model $p_\mathrm{pop}(\boldsymbol{\theta}|\boldsymbol{\Lambda})$ provides a prior for how probable those source properties are given the underlying population.
Since not all potential sources in the population model will be detected, we normalise by a selection function $\alpha_\mathcal{S}(\boldsymbol{\Lambda})$ for each source type. 
The exponential is a Poisson term accounting for the probability of detecting $N_\mathrm{det}^\mathcal{S}$ sources given a mean rate $\mathcal{R}_\mathcal{S}\alpha$~\citep{Mandel:2018mve, Vitale:2020aaz}.  

\subsection{Signal simulation}
\label{sec:signal-models}

For each \ac{EMRI}, we simulate a \ac{GW} signal to be observed by \ac{LISA}. 
We used the \texttt{Pn5AAK} waveform model for our analysis, implemented in the \texttt{FastEMRIWaveforms} package~\citep{Chua:2020stf, Katz:2021yft, Speri:2023jte, Chapman-Bird:2025xtd}. 
The waveforms are then modulated by the \ac{LISA} response function, and the detector noise is modelled using the analytic \texttt{SciRDv1} power spectral density~\citep{Babak:2021mhe}.

For each \ac{TDE} event, we simulate the survey response using the \texttt{SimulateOpticalTransient} module of \texttt{Redback}~\citep{Sarin:2023uqa}, convolving the theoretical light curve with the observing cadence, depth, and noise properties of \ac{ZTF} (bands $\{g,r,i\}$) and \ac{LSST} (bands $\{u,g,r,i,z,y\}$).
We restrict to the \texttt{tde\_fallback} model implemented in \texttt{Redback}~\citep{Sarin:2023uqa}. 
For each simulated observation, the photometric uncertainty is set directly to the survey's per-visit $5\sigma$ limiting magnitude for that pointing, as recorded in the pointing-history tables.
The observed flux is then a Gaussian realisation centred on the true source flux, with this limiting-magnitude uncertainty.

\subsection{Single event parameter estimation}
\label{sec:single-pe}

To estimate the hyperposterior $p(\boldsymbol{\Lambda}|\boldsymbol{d})$ in Eq.~\eqref{eq:hyperlikelihood_k}, we need to evaluate the event-level likelihood for the $i$-th event $p(\boldsymbol{d}|\boldsymbol{\theta}_i)$.
The \ac{EMRI} and \ac{TDE} parameters $\boldsymbol{\theta}_i$ can be extracted from the data streams of \ac{LISA}, \ac{ZTF}, and \ac{LSST} using event-level parameter-estimation methods.
For computational efficiency, we perform a Fisher matrix analysis in which the likelihood is approximated by a multivariate Gaussian~\citep[Chapter 7]{Maggiore:2007ulw}.
This approximation is valid in the high-\ac{SNR} regime, where the likelihood is well approximated by a Gaussian~\citep{Cutler:1994ys, Vallisneri:2007ev}.
To account for the influence of noise, we randomly offset the maximum likelihood point from the true value, self-consistently using a distribution derived from the Fisher matrix evaluated at the true values~\citep{Stevenson:2017dlk, Chapman-Bird:2022tvu, Singh:2026sad}.

We calculate Fisher matrices for \acp{EMRI} using the \texttt{StableEMRIFisher} package~\citep{kejriwal_2024_sef}.

For \acp{TDE}, the Fisher matrix is constructed from numerical derivatives of the light curve model with respect to each source parameter.
A \ac{TDE} observed in bands $\{X\}$ at times $\{t_n\}$, the Fisher matrix is
\begin{equation}
\label{eq:TDE_fisher}
    F_{ij}^{\mathrm{TDE}} 
    = \sum_{X} \sum_{k}
      \frac{1}{\sigma_m^2}
      \frac{\partial m_X(t_n;\boldsymbol{\theta})}{\partial \theta_i}
      \frac{\partial m_X(t_n;\boldsymbol{\theta})}{\partial \theta_j},
\end{equation}
where $m_X(t_n;\boldsymbol\theta)$ is the model's apparent magnitude in band $X$ (Table~\ref{tab:bands_cadence}) at time $t_n$ and $\sigma_m$ is the photometric magnitude uncertainty.
The derivatives in Eq.~\eqref{eq:TDE_fisher} are computed numerically using a two-sided finite-difference scheme, with step sizes adapted to whether each parameter enters the model linearly or logarithmically.

\subsection{Phenomenological population models}
\label{sec:population_distribution}

The framework described in Sec.~\ref{sec:astrophysical_setup} allows us to generate physically motivated, time‑dependent rates of \acp{EMRI} and \acp{TDE} across a population of galaxies.
This provides a forward model that predicts the distribution and evolution of \acp{EMRI} and \acp{TDE} across the simulated galaxy population. 
In reality, we do not know the true distribution of sources.
To connect the distribution and evolution of \acp{EMRI} and \acp{TDE} to observables, we introduce a phenomenological description for $p_\mathrm{pop}(\boldsymbol{\theta}|\boldsymbol{\Lambda})$.
This approach allows us to remain agnostic about the detailed microphysics of \ac{MBH} growth and evolution while retaining sufficient flexibility to capture distinct formation channels and evolutionary scenarios.

Following \citet{Singh:2026sad}, we consider two representative phenomenological population models, $\mathcal{A}$ and $\mathcal{B}$, which differ in complexity: their descriptions are presented in Table~\ref{tab:models}.

We infer the distributions for the \ac{MBH} masses, spins, eccentricity, plunge, and source redshift, while holding the other distributions fixed. 
For Model~$\mathcal{A}$ we adopt a modified Schechter-like distribution~\citep{Schechter:1976iz} for the \ac{MBH} mass, which better reflects the underlying distribution than the power-law distribution of Model~$\mathcal{B}$~\citep{Babak:2017tow, Barausse:2020mdt, LIGOScientific:2026ctl, Habouzit:2021bxi}. 
Both spin distributions are bimodal, with Model~$\mathcal{A}$ a bimodal Beta distribution~\citep{Wysocki:2018mpo} and Model~$\mathcal{B}$ a bimodal Gaussian distribution. 
In both models, we assume the primary peak to be $a_\bullet\sim0.3$ for \acp{MBH} $\lesssim10^6~M_\odot$ and $a_\bullet \sim0.7$ for \acp{MBH} $\gtrsim10^6~M_\odot$. 
We similarly parametrise $e_{\mathrm{pl}}$ as a bimodal Beta for Model~$\mathcal{A}$ and a bimodal Gaussian distribution for Model~$\mathcal{B}$.
We model the redshift distribution as the comoving-volume element times a power-law evolution with a fixed exponential high-redshift cut-off ($z_\mathrm{cut} = 0.9$),
\begin{equation}
    p(z\mid\kappa_z) \propto \frac{\mathrm{d}\mathcal{V}_\mathrm{c}}{\mathrm{d}z}(z)\,(1+z)^{\kappa_z}\,e^{-z/z_\mathrm{cut}},
    \label{eq:pz_kappa}
\end{equation}
for both models.

\begin{table*}
\centering
\footnotesize
\caption{Phenomenological population models considered.
Models $\mathcal{A}$ and $\mathcal{B}$ are shown along with the prior space for each parameter.
Each model corresponds to specific functional choices and hyperparameters, as detailed in Table~\ref{tab:generic_form}.}
\label{tab:models}
\begin{tabular}{lccccc}
\hline
Category & Parameter ${\theta}_j$ & Description & Model~$\mathcal{A}$ & Model~$\mathcal{B}$ & Range \\
\hline

\multirow{4}{*}{Shared}
& $\log_{10}(M_\bullet/M_\odot)$ & Source-frame log \ac{MBH} mass & Schechter & Power-law & [$4$, $7$]\\
& \ac{CO} mass ($M_\odot$) & \ac{CO} mass (\ac{sBH}/star) & Bimodal Gaussian & Bimodal Gaussian & \{$1$, $10$\}\\
& $a_\bullet$ & \ac{MBH} spin & Bimodal Beta & Bimodal Gaussian & [$0.01$, $0.99$]\\
& $z_{\mathrm{gal}}^{\mathrm{obs}}$ & Redshift & Eq.~\eqref{eq:pz_kappa} & Eq.~\eqref{eq:pz_kappa} & ($0$, $z_{\max}$]\\
\hline

\multirow{9}{*}{\ac{EMRI}}
& $e_{\mathrm{pl}}$ & Eccentricity at plunge & Bimodal Beta & Bimodal Gaussian & [$0.01$, $0.99$]\\
& $p_0$ & Initial semi-latus rectum & --- & --- & $p_0(T_\mathrm{plunge})$ \\
& $\cos\iota_0$ & Cosine inclination angle & Uniform & Uniform & $[-1,\,1]$ \\
& $\theta_\mathrm{S}$ & Polar sky location & Uniform & Uniform & ($0$, $\pi$) \\
& $\phi_\mathrm{S}$ & Azimuthal sky location & Uniform & Uniform & ($0$, $2\pi$) \\
& $\theta_\mathrm{K}$ & Polar spin orientation & Uniform & Uniform & ($0$, $\pi$) \\
& $\phi_\mathrm{K}$ & Azimuthal spin orientation & Uniform & Uniform & ($0$, $2\pi$) \\
& $\Phi_{\phi,0}$, $\Phi_{\theta,0}$, $\Phi_{r,0}$ & Initial orbital phases & Uniform & Uniform & [$0$, $2\pi$) \\
& $T_\mathrm{plunge}$ (yr) & Plunge time since observation start & Uniform & Uniform & [$1$, $6$] \\
\hline

\multirow{4}{*}{\ac{TDE}}
& $\beta_{\mathrm{orb}}$ & Penetration factor & Uniform & Uniform & [$1.001$, $30$] \\
& $\alpha_{\mathrm{visc}}$ & Viscosity parameter & Uniform & Uniform & [$0.01$, $0.1$] \\
& $\eta$ & Radiative feedback efficiency & Uniform & Uniform & [$10^{-3}$, $10^{-1}$] \\
\hline

\end{tabular}
\end{table*}

\subsection{Selection functions}
\label{sec:selection-function}

The detected \acp{EMRI} and \acp{TDE} represent a biased subpopulation of all signals. 
The selection function $\alpha_\mathcal{S}(\boldsymbol{\Lambda})$ quantifies the detectable fraction of the underlying population, enabling this bias to be accounted for~\citep{Mandel:2018mve, Vitale:2020aaz}.
We treat the detectability of \ac{EMRI} and \ac{TDE} solely as a function of \ac{SNR}.
It can be expressed in terms of the detection probability $P_{\mathrm{det}}(\boldsymbol{\theta})$ characterised by the source parameters $\boldsymbol{\theta}$, such that
\begin{equation}
\label{eq:sf_general}
    \alpha_\mathcal{S}(\boldsymbol{\Lambda})
    =
    \int \mathrm{d}\boldsymbol{\theta} \
    P^\mathcal{S}_{\mathrm{det}}(\boldsymbol{\theta})
    p_{\mathrm{pop}}(\boldsymbol{\theta}\mid\boldsymbol{\Lambda}).
\end{equation}
The detection probability $P_{\mathrm{det}}^\mathcal{S}(\boldsymbol{\theta})$ encodes the response of each instrument to a source with parameters $\boldsymbol{\theta}$, and it differs between the two messengers.

For \acp{EMRI}, the detection probability $P_{\mathrm{det}}^{\mathrm{EMRI}}(\boldsymbol{\theta})$ in Eq.~\eqref{eq:sf_general} is assumed to be~\citep[Chap.~7]{Maggiore:2007ulw}
\begin{equation}
\label{eq:Pdet_EMRI}
    P_{\mathrm{det}}^{\mathrm{EMRI}}(\boldsymbol{\theta}) = \mathcal{H}(\rho_n-\rho^\mathrm{EMRI}_\mathrm{t}),
\end{equation}
where $\rho_n$ is a noise-realised matched-filter \ac{SNR} and $\rho^\mathrm{EMRI}_\mathrm{t} = 20$ is the threshold \ac{SNR}~\citep[]{MockLISADataChallengeTaskForce:2009wir}.
We obtain $\rho_n^2$ by drawing a sample from a non-central $\chi^2$ distribution with two degrees of freedom and non-centrality parameter $\rho_{\mathrm{opt}}^2$~\citep[Chapter 7]{Maggiore:2007ulw}, where $\rho_{\mathrm{opt}}$ is the optimal \ac{SNR}~\citep{Chapman-Bird:2022tvu}. 

The selection function in Eq.~\eqref{eq:sf_general} can then be evaluated as a Monte Carlo sum 
\begin{equation}
    \label{eq:EMRI_SF}
    \alpha_{\mathrm{EMRI}}(\boldsymbol{\Lambda})  \simeq \frac{1}{N_\mathrm{t}}\sum_{\mathcal{G}=1}^{N_\mathrm{t}}{\overline{P^{\mathrm{EMRI}}_\mathrm{det}}}(\boldsymbol{\theta}_\mathcal{G}),
\end{equation}
where $\{\boldsymbol{\theta}_\mathcal{G}\}$ are sampled from the population distribution $p_\mathrm{pop}(\boldsymbol{\theta}|\boldsymbol{\Lambda})$ and $\overline{P^{\mathrm{EMRI}}_\mathrm{det}}=1-P(\rho_n>\rho_\mathrm{t}|\rho_\mathrm{opt})$ and $N_\mathrm{t}$ is the number of drawn samples.

The detectability of \acp{TDE} is defined through the response of the survey to a light curve characterised by $\boldsymbol{\theta}$.
The detection of a \ac{TDE} depends on a sufficient number of photometric detections each above the detection threshold of the survey~\citep{Bellm:2018cps, Ahumada:2024qpr}.
We therefore adopt an empirical, binary approach to model the detectability. 

For each \ac{TDE} event, we simulate the survey response, generating a realistic photometric time series by convolving the theoretical light curve $m_X(t;\boldsymbol{\theta})$ in band $X$ with the real observing cadence, depth, and noise properties of the survey; this removes the need to sample from an analytic noise distribution as for \acp{EMRI}, since the stochasticity due to photometric noise is already realised in each simulated observation.

\begin{figure}
    \centering
    \begin{subfigure}{0.5\textwidth}
        \includegraphics[width=0.8\linewidth]{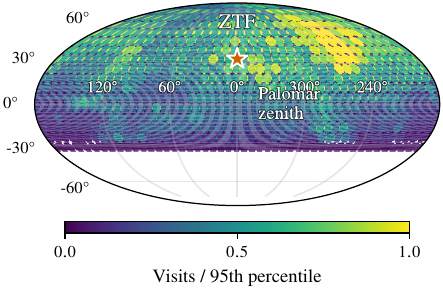}
        \caption{ZTF era}
        \label{fig:footprint-ztf}
    \end{subfigure}
    \hfill
    \begin{subfigure}{0.5\textwidth}
        \includegraphics[width=0.8\linewidth]{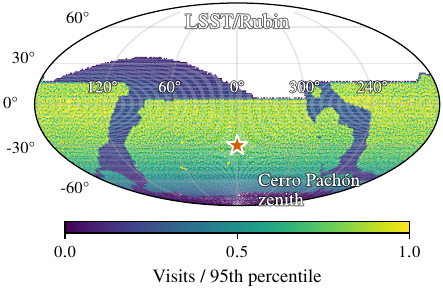}
        \caption{LSST/Rubin era}
        \label{fig:footprint-lsst}
    \end{subfigure}
    \caption{Sky footprint and pointing-density completeness of \ac{ZTF}~\citep{Bellm:2018cps, Masci:2018neq} (left) and LSST/Rubin~\citep{Bianco:2021ape} (right), built from \texttt{Redback}'s pointing-history tables~\citep{Sarin:2023uqa}. 
    Colour shows the number of recorded visits per sky cell; white regions are never observed, and stars mark the zenith declination of each observatory.}
    \label{fig:footprint}
\end{figure}

We consider two surveys, \ac{ZTF}, with bands $\{g, r, i\}$, and \ac{LSST}, with bands $\{u, g, r, i, z, y\}$.
Single-site ground-based surveys observe only the sky that rises above the local horizon over the course of a year~\citep{LSST:2008ijt, Bellm:2018cps}, a geometric limit independent of source brightness and distinct from the magnitude-limited selection within the footprint.
\ac{ZTF} (Palomar, $+33\fdg35$ latitude) executes its Northern Sky Survey over $\delta \gtrsim -31\degr$~\citep{Bellm:2018cps, Graham:2019qsw, Bellm:2018cps}, whereas \ac{LSST}/Rubin (Cerro Pach\'on, $-30\fdg24$) covers a complementary southern and equatorial $\sim$18,000--20,000\,deg$^2$~\citep{LSST:2008ijt, Bianco:2021ape}, excluding, for example, parts of the Galactic plane in the wide--fast--deep cadence~\citep{Bianco:2021ape}.
Detectability within the footprint is also non-uniform~\citep{Dekany:2020tyb}.
The net completeness rises with declination from $\sim$50\% near the southern edge ($\delta \sim -13\degr$) to $\sim$76\% at $\delta > 40\degr$~\citep{Chen:2020veh, Masci:2018neq}, reflecting visit cadence, airmass, and seasonal visibility.
The detectability of a simulated \ac{TDE} is therefore set by a hard geometric mask (the footprint) and a soft completeness gradient within it (the normalised density of recorded visits and, where relevant, the achieved limiting magnitude), both shown in Fig.~\ref{fig:footprint} with the zenith declination of each observatory marked.

For each simulated observation in band $X$ at time $t_i$, we obtain an observed (noise-realised) flux $\mathcal{F}_{X,t_i}$ and an associated photometric uncertainty $\sigma_{X,t_i}$, set by the survey's real per-visit limiting-magnitude noise at that pointing. 
The signal-to-noise ratio of that observation is then given as 
\begin{equation}
\label{eq:snr_definition_tde}
    \rho_{X,t_i} = \frac{\mathcal{F}_{X,t_i}}{\sigma_{X,t_i}}.
\end{equation}
We adopt a detection threshold of $\rho^\mathrm{TDE}_\mathrm{t} = 5$ for both survey following the $5\sigma$ limiting-depth definitions of ZTF~\citep{Bellm:2018cps, Masci:2018neq} and LSST~\citep{LSST:2008ijt, Bianco:2021ape}, requiring this to be reached in at least reached in at least $N_\mathrm{min} = 2$ bands.
The detection probability, therefore, follows
\begin{equation}
\label{eq:Pdet_TDE}
P_{\mathrm{det}}^{\mathrm{TDE}}(\boldsymbol{\theta}) 
        = \mathbb{I}\left[
        \exists\, X \in \{\text{survey bands}\} :
        \sum_{t_i} \mathcal{H}(\rho_{X,t_i} - \rho^\mathrm{TDE}_\mathrm{t}) \geq N_{\mathrm{min}}
        \right],
\end{equation}
where $t_i$ is the photometric observation time, $\{\text{survey bands}\}$ denotes $\{g,r,i\}$ for \ac{ZTF} or $\{u,g,r,i,z,y\}$ for \ac{LSST}, and the indicator $\mathbb{I}[\cdot]$ returns $1$ if the condition is satisfied and $0$ otherwise.
Each simulated light curve is followed for a finite window after disruption, $500~\mathrm{d}$ for \ac{ZTF} and $700~\mathrm{d}$ for \ac{LSST}, which sets the maximum baseline over which the $N_{\mathrm{min}}=2$ criterion can be satisfied. 
\texttt{Redback} samples each light curve at the recorded visit times at the sky position of the source, so the effective cadence varies between fields; the median revisit time in each band is given in Table~\ref{tab:bands_cadence}~\citep{Sarin:2023uqa}.

To calculate the selection function for \acp{TDE}, we adopt a similar approach as for \acp{EMRI} and perform a Monte Carlo sum.
\begin{equation}
\label{eq:sf_TDE}
    \alpha_{\mathrm{TDE}}(\boldsymbol{\Lambda}) 
    \simeq \frac{1}{N_{\mathrm{t}}} 
    \sum_{\mathcal{G}=1}^{N_{\mathrm{t}}} 
    P_{\mathrm{det}}^{\mathrm{TDE}}(\boldsymbol{\theta}_\mathcal{G}), 
\end{equation}
where $\{\boldsymbol{\theta}_\mathcal{G}\}$ are sampled from the population distribution $p_\mathrm{pop}(\boldsymbol\theta\mid\boldsymbol\Lambda)$, $P_{\mathrm{det}}^{\mathrm{TDE}}$ is the binary detection outcome defined in Eq.~\eqref{eq:Pdet_TDE} for a single noise realisation of the simulated light curve, and $N_\mathrm{t}$ is the number of drawn samples.

To achieve a percent-level accuracy, one needs to compute $P_\mathrm{{det}}^\mathcal{S}(\boldsymbol{\theta}_\mathcal{G})$ $\sim10^5$--$10^6$ times, which is prohibitively expensive considering that $\alpha\mathcal{S}(\boldsymbol{{\Lambda}})$ must be evaluated a similar number of times during a typical population inference.
To overcome this bottleneck in the selection function, we use \texttt{poplar}~\citep{chapman_bird_2025_17897846}, developed in \citet{Chapman-Bird:2022tvu} and further extended for \ac{EMRI} population inference in \citet{Singh:2025pzh}.
Two \ac{MLP} neural networks~\cite[Sec.\ 8.2.1]{Acquaviva:2023gkb} are used to emulate (i) the \ac{SNR} and (ii) the selection function, the latter trained directly on the outputs of the former. 
We extend \texttt{poplar} to \acp{TDE} by training a separate pair of \ac{SNR} and selection-function \acp{MLP} on the simulated \ac{ZTF} and \ac{LSST} detectability outcomes described above.
This framework provides orders-of-magnitude speed-ups in direct evaluations while enabling unbiased estimation of population parameters for both messengers.

For \acp{TDE}, the selection-function \ac{MLP} is a binary-classification network trained with a binary cross-entropy loss on $10^6$ simulated events per survey to predict a detection probability from the source parameters.
An event is counted as detected if this predicted probability exceeds a threshold of $0.5$.

\subsection{Inference setup}
\label{sec:inference-setup}

Having calculated all the necessary quantities to compute the hyperlikelihood, we employ nested sampling using \texttt{nessai}~\citep{nessai, Williams:2021qyt} to explore the hyperparameter space efficiently.
All hyperparameters are assigned flat (uniform) priors over the bounds (Table~\ref{tab:hyperprior_values}).
We run \texttt{nessai}'s \texttt{FlowSampler} with $n_\mathrm{live}=1000$ live points, a fixed random seed for reproducibility, and deploy parallel worker processes for likelihood evaluation.

\section{Population Inference Results}
\label{sec:population_inference_results}

We now apply the framework developed in Sec.~\ref{sec:Joint_Hierarchical_Inference} to the mock catalogues corresponding to the eight population scenarios listed in Table~\ref{tab:simulated_catalogue_scenario_summary}.
Each scenario is analysed using the two phenomenological models, Model~$\mathcal{A}$ and Model~$\mathcal{B}$ (Table~\ref{tab:models}), across eight dataset combinations, collectively denoted by $\boldsymbol{d}$.

In the following subsections, we present the constraints obtained from the individual datasets: \acp{EMRI} ($\boldsymbol{d}_{\mathrm{EMRI}}$), \acp{TDE} detected by \ac{ZTF} ($\boldsymbol{d}_{\mathrm{ZTF}}$), and \acp{TDE} detected by \ac{LSST} ($\boldsymbol{d}_{\mathrm{LSST}}$) in Sec.~\ref{subsec:shared_mass_redshift}. 
We then investigate how the constraints improve when \ac{TDE} catalogues are combined with the \ac{EMRI} observations in Sec.~\ref{subsec:improvement_EM}. 
To do so we analyse the combined catalogues consisting of \acp{EMRI} and \ac{ZTF}-\acp{TDE} ($\boldsymbol{d}_{\mathrm{EMRI+ZTF}}$), \acp{EMRI} and \ac{LSST}-\acp{TDE} ($\boldsymbol{d}_{\mathrm{EMRI+LSST}}$), and \acp{EMRI} together with the combined \ac{ZTF}+\ac{LSST}-\acp{TDE} catalogue ($\boldsymbol{d}_{\mathrm{EMRI+LSST+ZTF}}$).
In Sec.~\ref{sec:spin_ecc_unshared} we consider constrains on spin and eccentricity, which are only measured by \acp{EMRI} in our modelling. 
Finally, we quantify the relative contribution of each source to the joint inference in Sec.~\ref{subsec:joint_contribution}. 
The analysis therefore addresses three questions: (1) which population-level properties each channel constrains, and how tightly; (2) how the constraints improve as catalogues are combined; and (3) the contribution of each source to understanding the properties of \acp{MBH}.

Before examining the individual hyperparameter constraints, we first assess how well the chosen phenomenological population models match the underlying catalogues (Table~\ref{tab:simulated_catalogue_scenario_summary}).
We compare the catalogues with the corresponding \acp{PPD}, which quantifies the distribution of expected sources implied by the inferred population model. 
For population hyperparameters $\boldsymbol{\Lambda}$ and future source parameters $\boldsymbol{\tilde{\theta}}$, the \ac{PPD} is given by 
\begin{equation}
p(\tilde{\theta}|\boldsymbol{d}) = \int p_\mathrm{pop}(\tilde{\theta}\mid\boldsymbol{\Lambda}) p(\boldsymbol{\Lambda}\mid\boldsymbol{d}) \mathrm{d}\boldsymbol{\Lambda}, 
\end{equation}
where $p_\mathrm{pop}(\tilde{\theta}|\boldsymbol{\Lambda})$ describes the population model and $p(\boldsymbol{\Lambda}|\boldsymbol{d})$ is the hyperposterior inferred from the observed catalogue.

\begin{figure*}
    \includegraphics[width=0.9\textwidth]{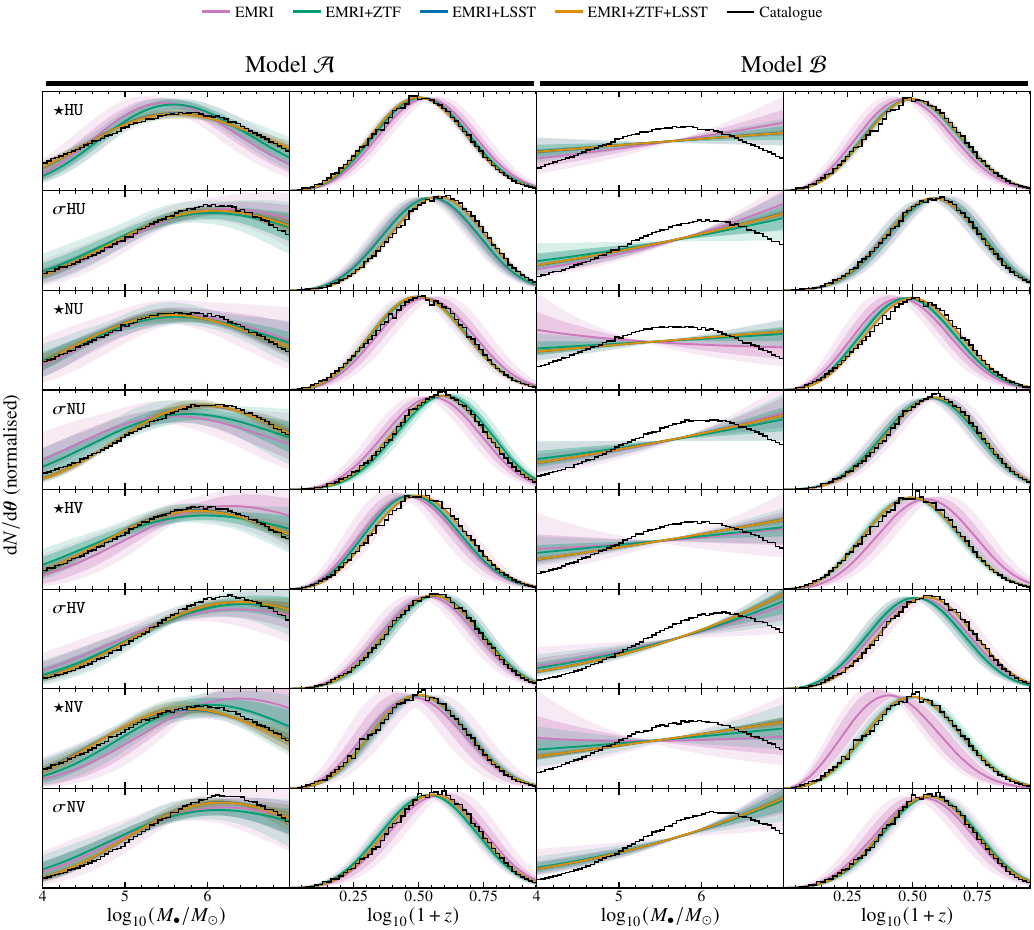}
    \caption{\Aclp{PPD} for the shared hyperparameters, the \ac{MBH} mass (left of each pair) and the redshift (right), for Model~$\mathcal{A}$ (left) and Model~$\mathcal{B}$ (right). 
    Coloured lines show the median and shaded bands the $90\%$ interval over posterior samples, for the \ac{EMRI}-only (pink), \ac{EMRI}+\ac{ZTF} (green), \ac{EMRI}+\ac{LSST} (blue) and \ac{EMRI}+\ac{ZTF}+\ac{LSST} (orange) catalogues. 
    The black lines show the eight underlying simulated population scenarios (Table~\ref{tab:models}).
    }
    \label{fig:PPD_shared}
\end{figure*}

\begin{figure*}
    \centering
    \includegraphics[width=0.9\textwidth]{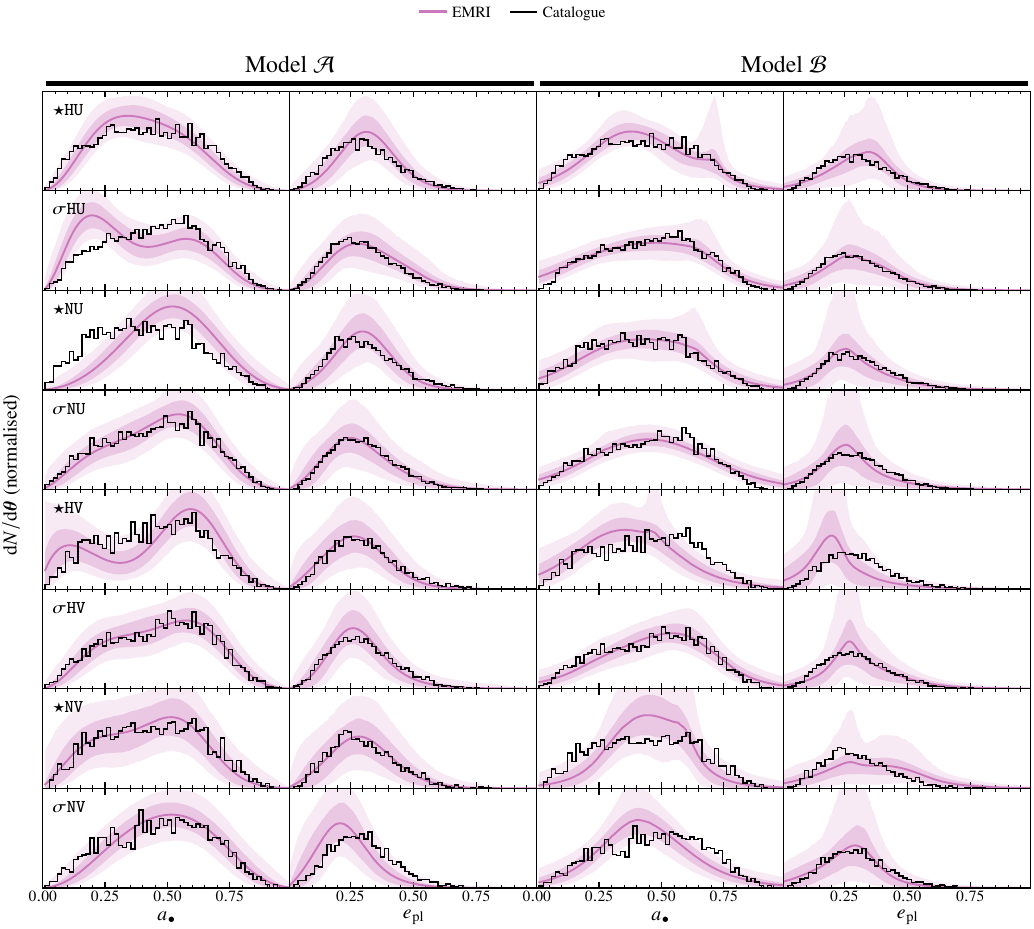}
    \caption{Similar to Fig.~\ref{fig:PPD_shared}, \acp{PPD} for the unshared parameters $a_\bullet$ (left of each pair) and $e_\mathrm{pl}$ (right), for Model~$\mathcal{A}$ (left) and Model~$\mathcal{B}$ (right).
    Coloured lines show the median and shaded bands the $90\%$ interval over posterior samples; we only show results for the \ac{EMRI}-only analysis as the \acp{TDE} add negligible information on these parameters. 
    The black lines show the eight underlying simulated population scenarios (Table~\ref{tab:models}).}
    \label{fig:PPD_unshared}
\end{figure*}

In Fig.~\ref{fig:PPD_shared} we plot the \acp{PPD} for the parameters common between \acp{EMRI} and \acp{TDE}, as well as the true underlying distributions from our simulations. 
In Fig.~\ref{fig:PPD_unshared} we similarly plot the parameters only measured by our \acp{EMRI} models. 
The underlying distributions represent the average over many realisations; the data used as input to the inference is only one realization drawn from this distribution, and so some variation between the \ac{PPD} and the underlying distribution is expected even for perfect choices of modelling functions.
In general, we find that there is good agreement between the inferred distributions, and the our target underlying distributions, with the simulation distribution lying within the uncertainties of the inference results. 
There are some cases, for example, the \ac{EMRI}-only redshift distribution for $\star\texttt{NV}$, where the small sample size means that the true distribution has not been as accurately reconstructed.  
However, agreement improves as more observations are added to the analysis. 
The worst recovered distribution appears to be the \ac{MBH} mass for Model~$\mathcal{B}$; as expected, the power-law form does not capture the true shape of the source distribution. 
Despite the inaccuracy of the Model~$\mathcal{B}$ mass distribution, the other parameters are still well recovered. 
This indicates that there are not strong correlations between parameters, such that a bias in one does not necessarily lead to biases in others.

\begin{figure}
    \centering    
    \includegraphics[width=\columnwidth]{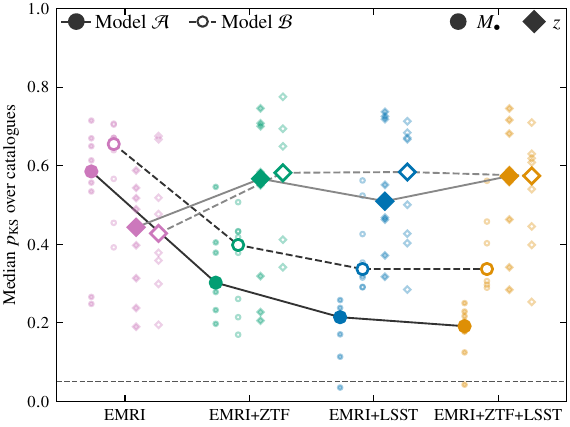}
    \caption{The \ac{KS} $p$-value for the \ac{PPD} against the underlying simulated population, for each dataset combination. 
    Large symbols are the median over the eight catalogues and small symbols the individual catalogues; circles show the \ac{MBH} mass and diamonds the redshift, for Model~$\mathcal{A}$ (filled) and Model~$\mathcal{B}$ (open).
    The dashed line marks $p_\mathrm{KS}=0.05$.
    }
    \label{fig:pv_by_dataset}
\end{figure}

For each scenario-catalogue combination, we evaluate the agreement between the \ac{PPD} and the simulations using the \ac{KS} statistic~\citep{Stephens:1974qsh}.
We compute the difference between their \aclp{CDF} $\mathrm{KS}(\boldsymbol{\Lambda}) = F_{\mathrm{PPD}}(\boldsymbol{\Lambda}) - F_{\mathrm{cat}}(\boldsymbol{\Lambda})$, where we draw a number of samples from each distribution to match the number of detections. 
If the fitted model accurately reproduces the catalogue, the \ac{KS} curve remains close to zero across the full parameter range.
From $\mathrm{KS}(\boldsymbol{\Lambda})$, we compute a $p$-value ($p_\mathrm{KS}$) for each model and dataset combination, and
Fig.~\ref{fig:pv_by_dataset} summarises this test, where we show the median $p_\mathrm{KS}$ over the eight simulated catalogues for Model~$\mathcal{A}$ and $\mathcal{B}$ for the shared parameters. 
Neither Model~$\mathcal{A}$ nor Model~$\mathcal{B}$ consistently show small $p$-values, indicating that given the number of observations, neither would be clearly identified by this test as a poor fit. 
These results indicate that caution is warranted when interpreting distributions reconstructed from these observations.
Even if a fitted distribution statistically represents the data well, it may not capture all the true features of the underlying source distribution. 
However, these tests indicate that both Model~$\mathcal{A}$ and Model~$\mathcal{B}$ yield results that could be representative of the state of inference following analysis of real \ac{EMRI} and \ac{TDE} observations.

\subsection{Constraints from individual datasets on mass and redshift}
\label{subsec:shared_mass_redshift}

The \ac{MBH} mass and the redshift distribution are the only hyperparameters to which both sources are sensitive, given the current state of modelling of the \ac{TDE} and \ac{EMRI} signals.
Since both observables are generated from the same underlying \ac{MBH} population, measurements of the mass function and its cosmic evolution can be combined from the two types of observations. 
The different datasets include independent observations and are subject to different selection effects.
\acp{TDE} provide large statistical samples that efficiently constrain the abundance of low-mass \acp{MBH}, while \acp{EMRI} observations provide an independent tracer of the same population through gravitational-wave measurements. 
Agreement between the inferred mass and redshift distributions from both sources would constitute an important validation of the underlying population model, whereas significant discrepancies could indicate missing astrophysical ingredients or limitations in current modelling assumptions.

For Model~$\mathcal{A}$, with $\boldsymbol{d}_{\mathrm{LSST}}$, the characteristic mass $x_c$ is constrained to a median $90\%$ width of $\aggPopAXcLsstWidMed~\mathrm{dex}$ and the shape parameter $\lambda_{x_c}$ to $\aggPopALamXcLsstWidMed~\mathrm{dex}$ over the eight scenarios.
The large \ac{LSST} sample therefore locates the characteristic mass to a fraction of a dex and only partially constrains the shape of the high-mass cut-off only partly.
The $\boldsymbol{d}_{\mathrm{ZTF}}$ and $\boldsymbol{d}_{\mathrm{EMRI}}$ constrain $x_c$ only to widths of $\aggPopAXcZtfWidMin$--$\aggPopAXcZtfWidMax~\mathrm{dex}$ and $\aggPopAXcEmriWidMin$--$\aggPopAXcEmriWidMax~\mathrm{dex}$, respectively, and leave $\lambda_{x_c}$ essentially unconstrained, with widths as wide as the prior range.
Neither channel on its own therefore resolves the shape of the turnover in Model~$\mathcal{A}$. 

In Model~$\mathcal{B}$, the sign of $\lambda_M$ indicates whether the population is weighted towards high-mass ($\lambda_M>0$) or low-mass ($\lambda_M<0$) \acp{MBH}. 
For the $\boldsymbol{d}_{\mathrm{LSST}}$ result, the power-law index $\lambda_M$ has a median width of $\aggPopBLamMLsstWidMed$ and is positive for every scenario, favouring a high mass distribution is $\lambda_M>0$. 
With the smaller \acp{EMRI} and \ac{ZTF}-\acp{TDE} samples, we find $\lambda_M > 0$ with probability $\aggPopBLamMSignAgreeEmriPctMed\%$ for the \acp{EMRI} and $\aggPopBLamMSignAgreeZtfPctMed\%$ for \ac{ZTF} (median over scenarios). 
Hence, the channels provide broad constraints on the shape of the mass distribution.

The redshift distribution is well constrained by the different datasets.  
With $\boldsymbol{d}_{\mathrm{LSST}}$, the power-law index $\kappa_z$ has $90\%$ widths of $\aggPopAKappaZLsstWidMin$--$\aggPopAKappaZLsstWidMax$ in Model~$\mathcal{A}$ and $\aggPopBKappaZLsstWidMin$--$\aggPopBKappaZLsstWidMax$ in Model~$\mathcal{B}$, about $2\%$ of the hyperprior range in both cases.
The $\boldsymbol{d}_{\mathrm{ZTF}}$ catalogues constrain it to $\aggPopAKappaZZtfWidMin$--$\aggPopAKappaZZtfWidMax$ and the $\boldsymbol{d}_{\mathrm{EMRI}}$ catalogues to $\aggPopAKappaZEmriWidMin$--$\aggPopAKappaZEmriWidMax$.
The redshift evolution is therefore measured mainly by the \acp{TDE}, and the few \acp{EMRI} constrain it only coarsely.
The $\boldsymbol{d}_{\mathrm{LSST}}$ median falls outside the $90\%$ interval of the $\boldsymbol{d}_{\mathrm{EMRI}}$ posterior only for $\star\,\texttt{NV}$ in Model~$\mathcal{B}$, and outside that of $\boldsymbol{d}_{\mathrm{ZTF}}$ only for $\sigma\,\texttt{NU}$ in Model~$\mathcal{A}$ and ($\sigma\,\texttt{HV}$) in Model~$\mathcal{B}$. 
Hence, the \acp{EMRI} and the \acp{TDE} agree on the redshift evolution in most scenarios.

\begin{figure*}
    \includegraphics[width=0.9\textwidth]{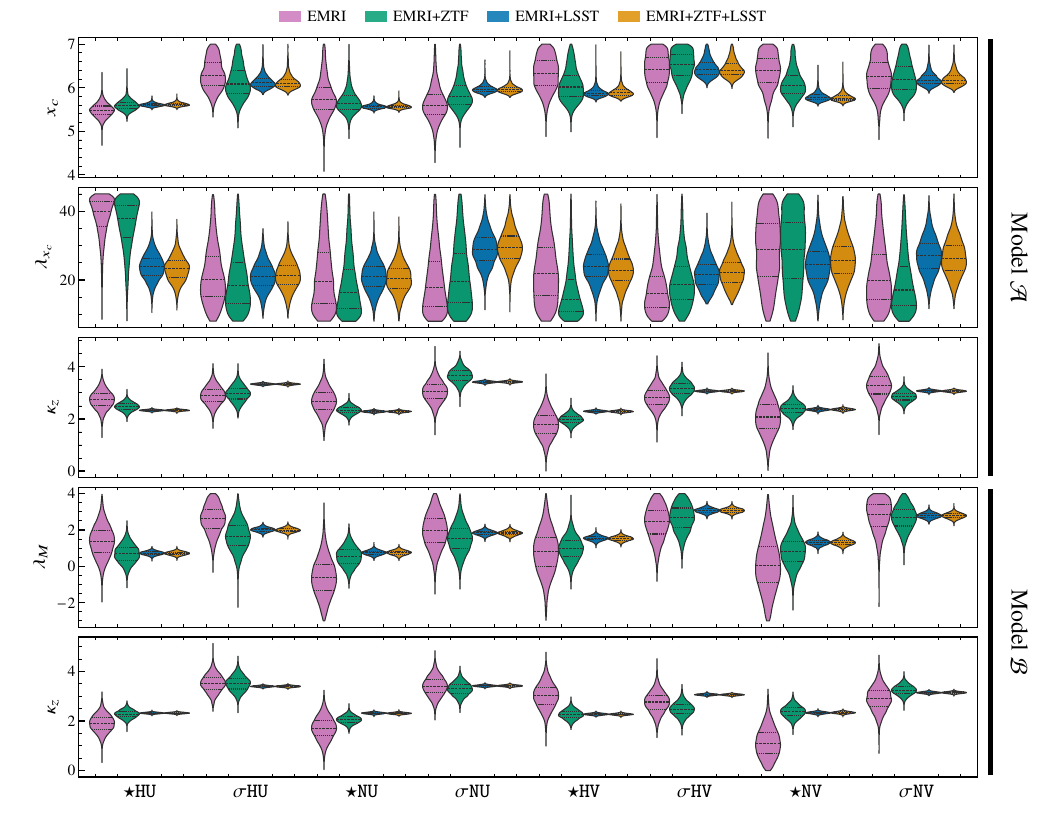}
    \caption{Marginal posteriors of the shared hyperparameters for each scenario-catalogue combination, for Model~$\mathcal{A}$ (top) and Model~$\mathcal{B}$ (bottom).}
    \label{fig:violon_shared_EMRI_TDE}
\end{figure*}

\subsection{Improvement in constraints by combining EMRIs and TDEs}
\label{subsec:improvement_EM}

\begin{figure}
\centering
\includegraphics[width=\columnwidth]{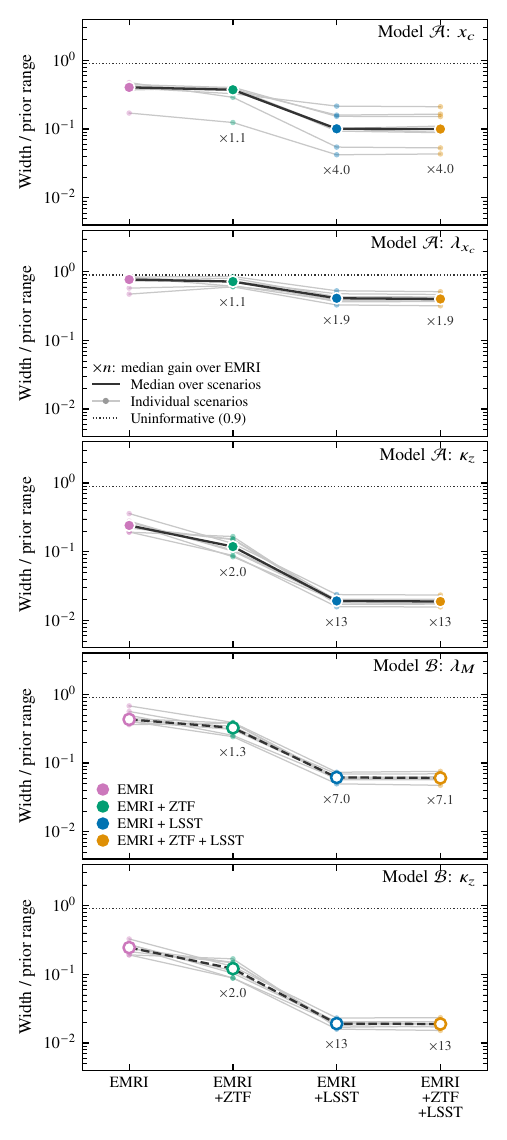}
\caption{Improvement of the constraints on the shared mass and redshift hyperparameters as \ac{TDE} catalogues are added to the \ac{EMRI} catalogue. Each panel shows the $90\%$ posterior width in units of the hyperprior range for the \ac{EMRI}-only, \ac{EMRI}+\ac{ZTF}, \ac{EMRI}+\ac{LSST} and \ac{EMRI}+\ac{ZTF}+\ac{LSST} catalogues (top row: Model~$\mathcal{A}$ filled-markers; bottom row: Model~$\mathcal{B}$, open-markers). 
Grey lines follow the individual scenarios and markers, joined by the black line, which shows the median across scenarios, annotated with the median gain relative to the \ac{EMRI}-only width. 
The dotted line marks the width of an uninformative posterior.
}
\label{fig:scaling_with_sources}
\end{figure}

As illustrated in Fig.~\ref{fig:PPD_shared}, combining datasets yields more precise results. 
We show the change in the $90\%$ widths of the shared mass and redshift hyperparameters of Model~$\mathcal{A}$ and Model~$\mathcal{B}$ \ac{TDE} catalogues when added to the \acp{EMRI} in Fig.~\ref{fig:violon_shared_EMRI_TDE} and Fig.~\ref{fig:scaling_with_sources}: in Fig.~\ref{fig:violon_shared_EMRI_TDE} we show the hyperposteriors for the population parameters, and in Fig.~\ref{fig:scaling_with_sources} we show how the widths of the hyperposteriors shrink with expanded data sets.
With \acp{EMRI} constituting only about $\aggPopAEmriFracZtfPctMin\%$--$\aggPopAEmriFracZtfPctMax\%$ of the \ac{EMRI}-\ac{TDE} catalogue, they provide only a coarse, order-of-magnitude anchor for the \ac{MBH} mass and redshift distributions, while the \ac{TDE} catalogues, by virtue of their larger event numbers, drive the bulk of constraint improvement.

With \acp{EMRI} alone, the median $90\%$ widths are about $\aggPopAXcEmriRelPctMed\%$ of the hyperprior range for $x_c$, $\aggPopALamXcEmriRelPctMed\%$ for $\lambda_{x_c}$ (Model~$\mathcal{A}$), $\aggPopBLamMEmriRelPctMed\%$ for $\lambda_M$ (Model~$\mathcal{B}$) and $\sim\aggPopAKappaZEmriRelPctMed\%$ for the redshift hyperparameter in both models
(Table~\ref{tab:hyperprior_values}, Fig.~\ref{fig:scaling_with_sources}).
The \ac{EMRI} catalogues therefore place a moderately informative constraint on the characteristic mass $x_c$, the mass slope $\lambda_M$ and the redshift evolution $\kappa_z$, and little information on the sharpness of the high-mass cut-off $\lambda_{x_c}$, whose posterior width spans most of the hyperprior range. 
Adding the \ac{ZTF}-\ac{TDE} catalogue improves these constraints modestly, most strongly for $\kappa_z$.
Relative to \acp{EMRI} alone, the median $90\%$ width decreases by factors of $\aggPopAXcEmriZtfGain$ for $x_c$, $\aggPopBLamMEmriZtfGain$ for $\lambda_M$, $\aggPopAKappaZEmriZtfGain$ in Model~$\mathcal{A}$ and $\aggPopBKappaZEmriZtfGain$ in Model~$\mathcal{B}$ for $\kappa_z$, and $\aggPopALamXcEmriZtfGain$ for $\lambda_{x_c}$ (Fig.~\ref{fig:scaling_with_sources}). 
An even larger improvement is obtained when the \ac{LSST}-\ac{TDE} catalogue is included. 
In this case, the median width is reduced by factors of $\aggPopAXcEmriLsstGain$ for $x_c$, $\aggPopBLamMEmriLsstGain$ for $\lambda_M$, $\aggPopAKappaZEmriLsstGain$ in Model~$\mathcal{A}$ and $\aggPopBKappaZEmriLsstGain$ in Model~$\mathcal{B}$ for $\kappa_z$, and $\aggPopALamXcEmriLsstGain$ for $\lambda_{x_c}$. Once the \ac{LSST} sample is included, however, the addition of \ac{ZTF} events produces only marginal further improvements, changing the median interval widths by at most $\aggBothZtfOnLsstMaxChangePct\%$. 
The strongest gains are obtained for the redshift evolution, followed by the mass slope $\lambda_M$.
For the characteristic mass scale, across the eight population scenarios, the $90\%$ credible interval width of $x_c$ for $\boldsymbol{d}_{\mathrm{EMRI+LSST}}$ lies between $\aggPopAXcEmriLsstWidMin$ and $\aggPopAXcEmriLsstWidMax\,\mathrm{dex}$ (Fig.~\ref{fig:scaling_with_sources}), demonstrating that the addition of \ac{TDE} observations substantially enhances the precision with which the shared population parameters can be constrained.

\subsection{Constraints on the spin and eccentricity}
\label{sec:spin_ecc_unshared}

The spin and eccentricity distributions are only measured by our \ac{EMRI} models. 
It is possible that the \ac{TDE} measurements could still improve constraints on these distributions if they helped break degeneracies. 
However, we find that hyperparameters are not strongly correlated, and there is negligible difference including the \acp{TDE}.
We therefore report \ac{EMRI}-only constraints (Fig.~\ref{fig:PPD_unshared}) for these distributions.

In the simulated catalogues, there are two channels with different eccentricity distributions; however, in the overall distributions, the two modes are not resolvable, and we find a single peak at $e_\mathrm{pl}\simeq0.2$--$0.3$ in all scenarios. 
This holds across all scenarios because the cliffhanger channel only becomes appreciable for $M_\bullet\lesssim3\times10^{5}~M_\odot$, and the \ac{LISA} detection threshold preferentially selects \acp{EMRI} from more massive hosts regardless of the assumed nucleation or occupation fraction.
Model~$\mathcal{A}$ correctly recovers the single mode by placing both Beta components at similar locations ($\simeq0.25$--$0.35$).
In Model~$\mathcal{B}$, the eccentricity \acp{PPD} are narrower and more sharply peaked than the catalogues in several scenarios ($\star \texttt{HV}$ and $\sigma \texttt{HV}$).
However, for $\star \texttt{NV}$, they develop a second mode near $e_\mathrm{pl}\simeq0.45$ to capture the higher-eccentricity tail. 
In all cases, we can correctly recover that the eccentricity distribution prefers low values and does not extend above $\sim0.75$.

The spin distribution is also bimodal by construction; in some catalogues the two modes can be resolved, while in others they merge together. 
A structure with multiple modes is difficult to accurately infer without a large sample size.
For Model~$\mathcal{A}$, two of the eight catalogue results show clear signs of bimodality ($\star \texttt{HV}$ and $\sigma \texttt{HU}$).
We recover both the modes in $\star \texttt{HV}$ and $\sigma \texttt{HU}$, though in $\sigma \texttt{HU}$ the \ac{PPD} overestimates the low-spin mode, and in $\star \texttt{HV}$, the \ac{PPD} overestimates the separation between modes.
In the remaining scenarios, Model~$\mathcal{A}$ has a single broad mode, except for $\star \texttt{NU}$, where the catalogue has a low-spin excess ($a_\bullet\lesssim0.3$) that the \ac{PPD} underestimates.
For Model~$\mathcal{B}$, the \ac{PPD} are predominantly unimodal in all eight scenarios.
However, some results also show narrow spikes in the $90\%$ band (e.g., near $a_\bullet\simeq0.7$ in $\star \texttt{HU}$), where there is a lower probability secondary mode. 
This behaviour may indicate that the bimodal Gaussian population model is less robust. 
For both Model~$\mathcal{A}$ and Model~$\mathcal{B}$, the finite number of detections limits how accurately the underlying distribution may be reconstructed.

\subsection{Contribution of each channel to the joint analysis}
\label{subsec:joint_contribution}

We have seen that the high number of \acp{TDE} makes them valuable for population inferences, and that \acp{EMRI} provide information that \acp{TDE} do not. 
Here, we evaluate how much information we gain from the different observations.

Adding more observations improves the precision of inferences. 
However, not every observation contributes equally.
The \acp{EMRI} make up only $\aggPopAEmriFracZtfPctMin\%$ to $\aggPopAEmriFracZtfPctMax\%$ of a joint \ac{EMRI}+\ac{ZTF} catalogue, yet adding the \acp{EMRI} shifts the posterior medians of the shared hyperparameters by up to $\aggBothJointShiftStdZtfMax$ standard deviations of the \ac{ZTF}-only posterior and narrows them by up to $\aggBothJointOverTdeWidthZtfMaxDevPct\%$ (Fig.~\ref{fig:violon_shared_EMRI_TDE} and Fig.~\ref{fig:scaling_with_sources}).
This signifies that the \acp{EMRI} carry disproportionate information per observation. 
With \ac{LSST}, when \acp{EMRI} make up at most $\aggPopAEmriFracLsstPctMax\%$ of detections, the joint and \ac{TDE}-only posteriors largely agree, with width ratios of $\aggBothJointOverTdeWidthLsstMin$ to $\aggBothJointOverTdeWidthLsstMax$ and median shifts of at most $\aggBothJointShiftStdLsstMax$ standard deviations.
The only appreciable change is in $\star\,\texttt{HU}$ of Model~$\mathcal{A}$, where the \acp{EMRI} narrows $x_c$ by about a quarter and moves $\lambda_{x_c}$ by about one standard deviation.
Hence, \acp{EMRI} are valuable, but rare compared to \acp{TDE}.

\begin{figure}
\centering
\includegraphics[width=\columnwidth]{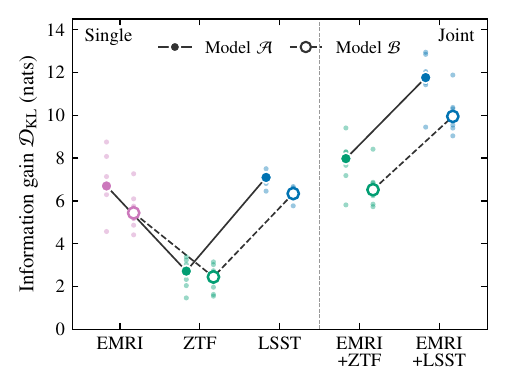}
\caption{Information gain $\mathcal{D}_\mathrm{KL}$ (nats) from hyperprior to hyperposterior for each catalogue. 
These are shown for Model~$\mathcal{A}$ (filled circles, solid lines) and Model~$\mathcal{B}$ (hollow circles, dashed lines).}
\label{fig:nats_info}
\end{figure}

To quantify how much we learn from the observations, we use he Kulback--Leibler divergence~\citep{Kullback:1951zyt} between the hyperprior and hyperposterior $\mathcal{D}_\mathrm{KL}$. 
This measures the information gained from the data. 
In Fig..~\ref{fig:nats_info}, we show the calculated $\mathcal{D}_\mathrm{KL}$ for different datasets.  
The \acp{EMRI} alone provide $\aggBothInfoEmriMin$--$\aggBothInfoEmriMax~\mathrm{nat}$; the \ac{ZTF} \acp{TDE} $\aggBothInfoZtfMin$--$\aggBothInfoZtfMax~\mathrm{nat}$, and the \ac{LSST} \acp{TDE} $\aggBothInfoLsstMin$--$\aggBothInfoLsstMax~\mathrm{nat}$.
Adding \acp{TDE} to an \ac{EMRI} catalogue contributes a further $\aggBothInfoEmriZtfMinusEmriMin$--$\aggBothInfoEmriZtfMinusEmriMax~\mathrm{nat}$ for \ac{ZTF} and $\aggBothInfoEmriLsstMinusEmriMin$--$\aggBothInfoEmriLsstMinusEmriMax~\mathrm{nat}$ for \ac{LSST}, whereas adding \acp{EMRI} to a \ac{TDE} catalogue contributes $\aggBothInfoEmriMinusTdeMin$--$\aggBothInfoEmriMinusTdeMax~\mathrm{nat}$.
The \ac{EMRI} information exceeds that of the \ac{ZTF} \acp{TDE} in every scenario, and is comparable to the \ac{LSST} \acp{TDE}: while the \acp{TDE} are more numerous, they do not constrain the spin and eccentricity hyperparameters.
The information gain quantifies how the two channels are complementary, showing that we do learn more from the combined dataset than from either observation type alone.

\section{Discussion and Conclusions}
\label{sec:discussion_conclusion}

%
Our results illustrate the complementary roles of \acp{TDE} and \acp{EMRI} as probes of the low-mass \ac{MBH} population.
Although \acp{TDE} and \acp{EMRI} are correlated via the nuclear dynamics of the underlying \ac{MBH} population, they exhibit different sensitivities to the astrophysical processes that shape nuclear environments.
Consequently, the combination of these observables offers a promising avenue for constraining both \ac{MBH} demographics and the dynamical evolution of galactic nuclei.

Our results support three main conclusions:
\begin{enumerate}
    \item \textbf{Detection rates can distinguish \ac{MBH} formation physics} (Fig.~\ref{fig:merger_model_impact_oleary} and \ref{fig:merger_model_impact_mundy_duan}) -- 
    Assuming our astrophysical models, we can already use detection rates to constrain \ac{MBH} physics. 
    Combinations of a high \ac{MBH} occupation fraction, efficient \ac{NSC} formation, and scaling relations that favour an enhanced low-mass \ac{MBH} population yield \ac{TDE} rates that \emph{exceed} those inferred from present-day surveys.
    The observed count favours scenarios in which fewer low-mass galaxies host both an \ac{MBH} and an \ac{NSC}, either through a lower nucleation fraction or an occupation fraction that falls towards low masses, combined with the $M_\bullet$--$\sigma$ relation.
    The dependence on both the occupation fraction and the \ac{NSC} nucleation fraction implies that \ac{TDE} samples can, in principle, discriminate between competing scenarios for the assembly of galactic nuclei.
    Larger samples are therefore likely to improve the constraints on the low-mass \ac{MBH} mass function and on the processes governing \ac{MBH} seeding and growth.
    Uncertainties in survey completeness and \ac{TDE} emission physics prevent definitive exclusions, but the existing \ac{TDE} catalogues already constrain regions of the parameter space that describes \ac{MBH} demographics.
    
    \item \textbf{The rate of \acp{TDE} places a limit on detectable \ac{LISA}-\acp{EMRI}} (Fig.~\ref{fig:emri_to_tde}) -- 
    Since both \acp{EMRI} and \acp{TDE} arise from the same \ac{NSC} dynamics, the scenarios allowed by the \ac{TDE} rate also bound the detectable \acp{EMRI}. 
    Across all the formation models we consider, the ratio remains consistent within a factor of a few. 
    Although the absolute rates of both phenomena remain subject to substantial astrophysical uncertainty, their ratio is a potentially informative observable because both populations originate from the same nuclear stellar environments. 
    Our study considers only specific formation pathways and neglects other potentially important pathways (e.g., partial and repeating \acp{TDE}, and wet \acp{EMRI}); hence, further study is needed to determine whether similar results hold more generally. 
    The \ac{EMRI}-to-\ac{TDE} ratio reflects the competition between disruption and inspiral pathways and is therefore sensitive to the efficiency of relaxation processes, the compact-object population, and the structure of the loss cone.
    Measurements of this ratio with future \ac{GW} and electromagnetic observations would provide a consistency check on models of nuclear stellar dynamics and \ac{EMRI} formation.
    
    \item \textbf{Both \acp{TDE} and \acp{EMRI} provide valuable information on the \ac{MBH} population} (Fig.~\ref{fig:scaling_with_sources} and Fig.~\ref{fig:nats_info}) -- 
    \acp{TDE} will provide constraints on the \ac{MBH} mass distribution prior to the launch of \ac{LISA}, but the precision measurements provided by \acp{EMRI} still add valuable information. 
    \acp{EMRI} make up only $\aggPopAEmriFracZtfPctMin$--$\aggPopAEmriFracZtfPctMax\%$ (median $\aggPopAEmriFracZtfPctMed\%$) of the detected \ac{EMRI}+\ac{ZTF} catalogue, yet the \ac{EMRI} catalogue alone yields $\aggBothInfoEmriMin$--$\aggBothInfoEmriMax~\mathrm{nat}$ of information about the \ac{MBH} population, more than the $\aggBothInfoZtfMin$--$\aggBothInfoZtfMax~\mathrm{nat}$ of the entire \ac{ZTF} sample.
    Most of this information concerns the spin and eccentricity distributions, which \acp{TDE} cannot access.
    With only a few \acp{EMRI}, these distributions are recovered in their broad shape, but not their finer structure, and the result depends on the population model (Fig.~\ref{fig:PPD_unshared}).
    The shared mass and redshift distributions are instead best constrained by the sheer number of \ac{LSST}-\acp{TDE} (Fig.~\ref{fig:violon_shared_EMRI_TDE}).

\end{enumerate}

By the time \acp{EMRI} are first detected, \ac{LSST} is expected to have detected thousands of \acp{TDE}~\citep{Karmen:2026ixp}, providing precise measurements of the abundance and mass distribution of low-mass \acp{MBH}. 
\acp{EMRI} will then contribute information that is inaccessible to \acp{TDE}, including the spin distribution of \acp{MBH}, the orbital properties of captured compact objects, and an independent probe of \ac{MBH} masses through their exceptionally precise \ac{GW} measurements. 
These mass measurements will provide a valuable cross-check of \ac{TDE}-based mass estimates, which depend on modelling electromagnetic observables such as light curves~\citep{Mockler:2018xne}. 

Given the potential importance of \ac{TDE} and \ac{EMRI} observations for uncovering properties of \acp{MBH}, it would be valuable to continue improving the modelling of their populations. 
For example, by simulating \acp{NSC} containing stellar components of a range of masses, and tracking relaxation processes as a function of time. It is also important to have accurate signal models (more realistic light curves and more complete \ac{GW} waveforms) so that source properties can be extracted from observations and selection effects can be calculated accurately. 
The statistical uncertainties associated with population inferences decrease as the number of observations increases, but systematic uncertainties depend on accurate source modelling.

This study has illustrated how combining two distinct observational tracers of the same underlying population within a common inference framework can yield better results than using either alone. 
Future work may consider whether additional observables could further enhance results. 
For example, how \ac{LISA} observations of merging \ac{MBH} binaries~\citep{Klein:2015hvg,LISA:2024hlh} can be used to constrain the \ac{MBH} population~\citep{Sesana:2010wy,Toubiana:2021iuw,Langen:2024ygz}. 
This may be more challenging, as while the \acp{TDE} and \acp{EMRI} considered here share a common formation pathway, other observations may probe different subpopulations of \acp{MBH} and their host environments. 
Hence, it would be necessary to consider \emph{evolutionary} selection effects in addition to observational selection effects~\citep[cf.][]{Liotine:2022vwq}. 
Additional information would be especially valuable if it could constrain properties such as the \ac{MBH} spin, which are only measured by the \acp{EMRI} in our modelling, and so are limited by the relatively small number of observations. 
By considering a range of observations, we may obtain a more complete picture of the astrophysics of \acp{MBH}.

\section*{Acknowledgements}
SS acknowledges support from the University of Glasgow.
CPLB acknowledges support from the UKSA grant UKRI972 and STFC grant UKRI2487. 
MV acknowledges funding from the French National Research Agency (grant ANR-21-CE31-0026, project MBH\_waves) and from the Centre National d’Etudes Spatiales (CNES).  
LB acknowledges the financial support provided under the European Union’s H2020 ERC Consolidator Grant “Binary Massive Black Hole Astrophysics” (B Massive, Grant Agreement: 818691) and the European Union Advanced Grant “PINGU” (Grant Agreement: 101142079).

\section*{Software}
\texttt{NumPy}~\citep{Harris:2020xlr},
\texttt{SciPy}~\citep{Virtanen:2019joe},
\texttt{Matplotlib}~\citep{Hunter:2007ouj},
\texttt{PyTorch}~\citep{Paszke:2019xhz},
\texttt{nessai}~\citep{nessai},
\texttt{StableEMRIFisher}~\citep{kejriwal_2024_sef},
\texttt{poplar}~\citep{chapman_bird_2025_17897846},
\texttt{FastEMRIWaveforms}~\citep{chapman_bird_2025_15624459},
\texttt{fastlisaresponse}~\citep{Katz:2022yqe},
\texttt{Redback}~\citep{Sarin:2023uqa},

\section*{Data Availability}
\label{sec:data_availability}

All data used in this work, comprising the simulated galaxy, \ac{EMRI} and \ac{TDE} catalogues, the single-event posterior samples, the hyperposterior samples of all inference runs and the figures, and scripts to generate the plots using \href{https://github.com/SSingh087/MBHpopulation/}{MBHpopulation} presented in this paper, are available on \href{https://doi.org/10.5281/zenodo.24393291}{Zenodo}.
The code used to generate the catalogues and perform hierarchical inference is available on \href{https://github.com/SSingh087/MBHpopulation/tree/main/MBH_population_from_EMRI_TDE}{GitHub}.

\appendix

\section{Validation for merger model}
\label{sec:merger_model_validation}

\begin{figure*}
    \centering
    \includegraphics[width=0.85\textwidth]{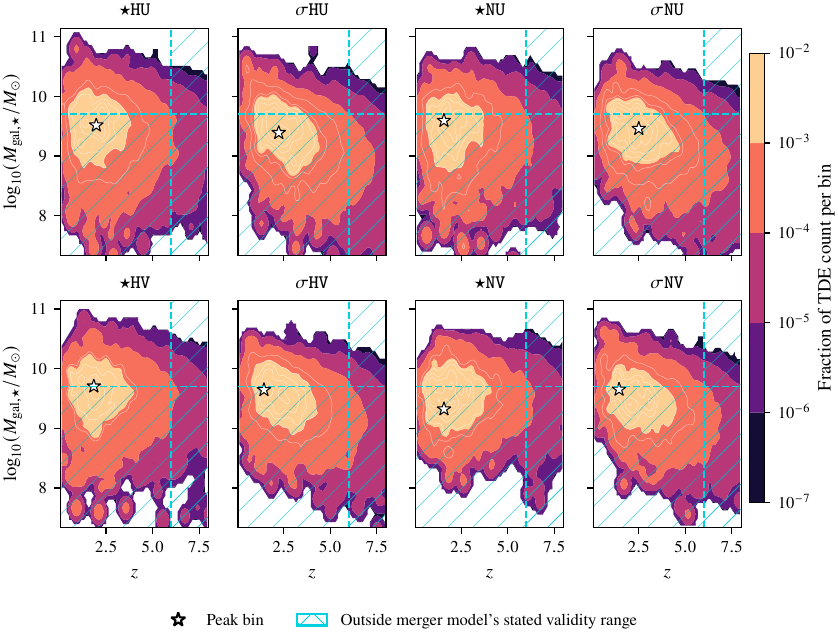}
    \caption{Simulated \ac{TDE} production weight (after nucleation, \ac{MBH} occupation, and the active-cusp cut) binned on a $(z,\log_{10}M_\star)$ grid for each of the eight population scenarios (Table~\ref{tab:simulated_catalogue_scenario_summary}), under the fiducial \citet{OLeary:2020edl} merger-rate model. 
    The star marks the smoothed peak-weight bin in each panel; the dashed red lines and shaded regions mark the stated validity bounds ($z\leq6$, $M_\star\geq10^{9.7}~M_\odot$) of \citet{OLeary:2020edl}, so that the fraction of each panel outside the shading is the fraction of the simulated production for which the merger rate is applied within its measured domain.}
    \label{fig:merger_validity_emri}
\end{figure*}

\begin{figure*}
    \centering
    \includegraphics[width=0.85\textwidth]{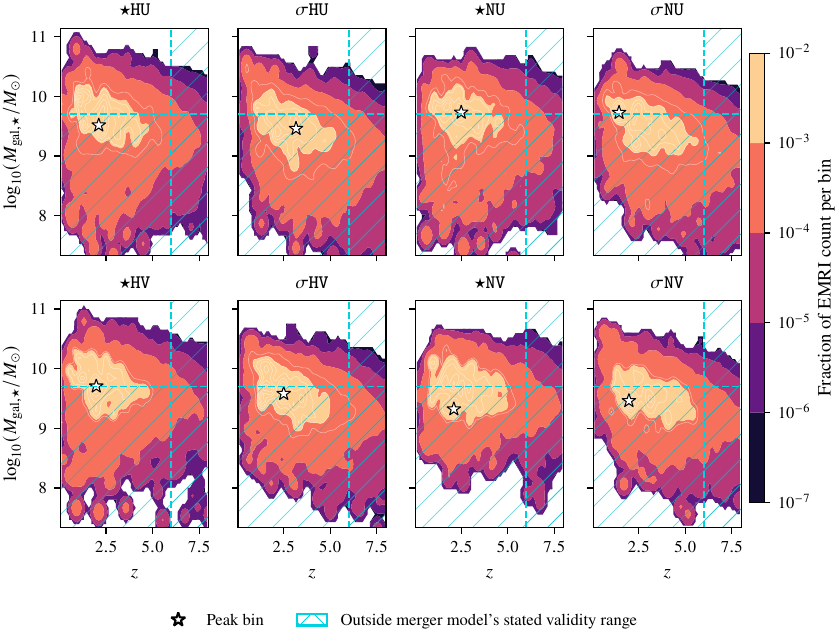}
    \caption{Simulated \ac{EMRI} production weight (after nucleation, \ac{MBH} occupation, and the active-cusp cut) binned on a $(z,\log_{10}M_\star)$ grid for each of the eight population scenarios (Table~\ref{tab:simulated_catalogue_scenario_summary}), under the fiducial \citet{OLeary:2020edl} merger-rate model.
    As in Fig.~\ref{fig:merger_validity_emri}, but for \acp{EMRI}.
    The star marks the smoothed peak-weight bin in each panel; the dashed red lines and shaded regions mark the stated validity bounds ($z\leq6$, $M_\star\geq10^{9.7}~M_\odot$) of \citet{OLeary:2020edl}, so that the fraction of each panel outside the shading is the fraction of the simulated production for which the merger rate is applied within its measured domain.}
    \label{fig:merger_validity_tde}
\end{figure*}

We quantified the impact of the choice of merger-rate model in two complementary ways.
Each of the merger models presented in Table~\ref{tab:merger_models} involves extrapolation since the galaxy mass and the redshift range do not cover the domain of our simulated galaxy catalogue entirely.
We show the region of validity of the merger models in Fig.~\ref{fig:merger_validity_emri} and Fig.~\ref{fig:merger_validity_tde}, where this extrapolation is explicit for each of the eight population scenarios (Table.~\ref{tab:simulated_catalogue_scenario_summary})
Additionally, we studied the impact of galaxy--galaxy merger models on the number of \ac{EMRI} and \ac{TDE} sources produced (Fig.~\ref{fig:merger_model_impact_oleary}).
In Fig.~\ref{fig:merger_validity_emri} and Fig.~\ref{fig:merger_validity_tde}, we bin the per-galaxy \ac{EMRI} weighted onto a two-dimensional $(z,\log_{10}M_\star)$ grid, after nucleation, \ac{MBH} occupation, and the merger-rate-driven active-cusp cut described in Sec.~\ref{sec:astrophysical_setup}.
We then overlay the redshift and mass validity bounds ($z\lesssim6$, $M_\star\geq10^{9.7}~M_\odot$) from \citet{OLeary:2020edl}  as dashed red lines, with the excluded region shaded.

Given the \citet{OLeary:2020edl} redshift and mass ranges, and the results' closeness to current \ac{TDE} predictions~\citep{vanVelzen:2020cwu, Yao:2023rbr, Hammerstein:2022wia}, we choose it as the base galaxy--galaxy merger model for our analysis.
Nonetheless, even the mass floor of \citet{OLeary:2020edl} sits roughly three decades above the lower edge of our simulated stellar-mass range, so the merger rate applied to the numerous low-mass galaxies in our catalogue remains an extrapolation beyond where any of the three literature fits was measured.
Closing this gap would require a mass-dependent merger-rate calibration extending to $M_\star\sim10^{8}~M_\odot$.

\begin{figure}
    \centering
    \includegraphics[width=\linewidth]{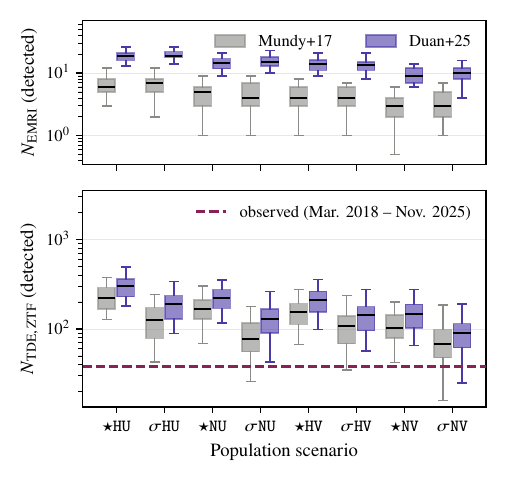}
    \caption{
    Similar to Fig.~\ref{fig:merger_model_impact_oleary}, the impact of the merger-rate model choice on the detected \ac{EMRI} ({top}) and \ac{TDE}-\ac{ZTF} ({bottom}) counts, for each of the eight population scenarios (Table~\ref{tab:simulated_catalogue_scenario_summary}), for $100$ realisations per scenario for the \citet{Mundy:2015qgi} (grey) and \citet{Duan:2025lqz} (grey) merger models. 
    Boxes show the interquartile range, with the median indicated by the horizontal black line; whiskers extend to the 5th and 95th percentiles.
    Detections are shown for a $6~\mathrm{yr}$ \ac{LISA} mission for \acp{EMRI} and for $8.75~\mathrm{yr}$ \ac{ZTF}-\acp{TDE} observations.
    The dashed line shows  confirmed \ac{ZTF}-\acp{TDE} detections~\citep{vanVelzen:2020cwu, Hammerstein:2022wia, Yao:2023rbr}.}
    \label{fig:merger_model_impact_mundy_duan}
\end{figure}

\section{Hyperpriors}
\label{app:hyperpriors}

All hyperparameters of Models~$\mathcal{A}$ and~$\mathcal{B}$ are assigned uniform hyperpriors (Sec.~\ref{sec:inference-setup}), with the bounds listed in Table~\ref{tab:hyperprior_values}.
The precision quoted in Sec.~\ref{sec:population_inference_results} is the $90\%$ posterior width relative to these ranges.

\begin{table}
\centering
\footnotesize
\caption{
    Hyperprior ranges for the population models considered in this work.
    Columns correspond to the population models $\mathcal{A}$ and $\mathcal{B}$
    For each hyperparameter $\Lambda_j \in \boldsymbol{\Lambda}$, the hyperprior $\pi(\Lambda_j)$ is set as a uniform distribution with the range $\Lambda_j \in [\Lambda_{\min}, \Lambda_{\max}]$ specified.
}
\label{tab:hyperprior_values}
\begin{tabular}{lcc}
\hline
Parameters $\boldsymbol{\theta}$ & $\mathcal{A}$ & $\mathcal{B}$ \\
\hline

\multirow{2}{*}{MBH mass}
& $x_c \in [\priorPopAXcLo,~\priorPopAXcHi]$
& \multirow{2}{*}{$\lambda_M \in [\priorPopBLamMLo,~\priorPopBLamMHi]$}
\\
& $\lambda_{{x_c}} \in [\priorPopALamXcLo,~\priorPopALamXcHi]$
&
\\
\hline

\multirow{5}{*}{Spin}
& $\alpha_{1,a} \in [\priorPopAAlphaOneSpinLo,~\priorPopAAlphaOneSpinHi]$
& $\mu_{1,a} \in [\priorPopBMuOneSpinLo,~\priorPopBMuOneSpinHi]$
\\
& $\alpha_{2,a} \in [\priorPopAAlphaTwoSpinLo,~\priorPopAAlphaTwoSpinHi]$
& $\sigma_{1,a} \in [\priorPopBSigmaOneSpinLo,~\priorPopBSigmaOneSpinHi]$
\\
& $\beta_{1,a} \in [\priorPopABetaOneSpinLo,~\priorPopABetaOneSpinHi]$
& $\mu_{2,a} \in [\priorPopBMuTwoSpinLo,~\priorPopBMuTwoSpinHi]$
\\
& $\beta_{2,a} \in [\priorPopABetaTwoSpinLo,~\priorPopABetaTwoSpinHi]$
& $\sigma_{2,a} \in [\priorPopBSigmaTwoSpinLo,~\priorPopBSigmaTwoSpinHi]$
\\
& $w_a \in [\priorPopAWeightSpinLo,~\priorPopAWeightSpinHi]$
& $w_a \in [\priorPopBWeightSpinLo,~\priorPopBWeightSpinHi]$
\\
\hline

\multirow{5}{*}{Eccentricity ($e_{\mathrm{pl}}$)}
& $\alpha_{1,e} \in [\priorPopAAlphaOneEccLo,~\priorPopAAlphaOneEccHi]$
& $\mu_{1,e} \in [\priorPopBMuOneEccLo,~\priorPopBMuOneEccHi]$
\\
& $\alpha_{2,e} \in [\priorPopAAlphaTwoEccLo,~\priorPopAAlphaTwoEccHi]$
& $\sigma_{1,e} \in [\priorPopBSigmaOneEccLo,~\priorPopBSigmaOneEccHi]$
\\
& $\beta_{1,e} \in [\priorPopABetaOneEccLo,~\priorPopABetaOneEccHi]$
& $\mu_{2,e} \in [\priorPopBMuTwoEccLo,~\priorPopBMuTwoEccHi]$
\\
& $\beta_{2,e} \in [\priorPopABetaTwoEccLo,~\priorPopABetaTwoEccHi]$
& $\sigma_{2,e} \in [\priorPopBSigmaTwoEccLo,~\priorPopBSigmaTwoEccHi]$
\\
& $w_e \in [\priorPopAWeightEccLo,~\priorPopAWeightEccHi]$
& $w_e \in [\priorPopBWeightEccLo,~\priorPopBWeightEccHi]$
\\
\hline

Redshift
& $\kappa_z \in [\priorPopAKappaZLo,~\priorPopAKappaZHi]$
& $\kappa_z \in [\priorPopBKappaZLo,~\priorPopBKappaZHi]$
\\
\hline
\end{tabular}
\end{table}

\section{TDE light-curve and survey settings}
\label{app:tde_settings}

This appendix lists the settings used to simulate the \ac{TDE} light curves and their detection (Sec.~\ref{sec:signal-models} and Sec.~\ref{sec:selection-function}).
Producing a synthetic light curve for a given survey band requires two ingredients.
First, the spectral energy distribution predicted by the model must be evaluated at the effective frequency $\nu_{\mathrm{eff}}$ of that band $X$.
This is what converts the bolometric light curve into the magnitude $m_X$ in that band, which is then compared with the survey limiting magnitude $m_{\mathrm{lim},X}$ to determine whether the event is detected.
Second, the cadence at which the band is sampled must be specified, since this sets how finely the rise and decay of the light curve are resolved and therefore what information can be recovered from it.
Table~\ref{tab:bands_cadence} gives, for each survey band, the effective frequency $\nu_{\mathrm{eff}}$ and wavelength $\lambda_{\mathrm{eff}}$ at which the spectral energy distribution is evaluated from \texttt{Redback}~\citep{Sarin:2023uqa} and the median revisit time between observations in that band, taken from the recorded pointing histories of \ac{ZTF}~\citep{Bellm:2018cps, Masci:2018neq} and \ac{LSST}~\citep{Bianco:2021ape}.
Table~\ref{tab:tde_lc_priors} lists the parameters of the \texttt{tde\_fallback} model used to generate the simulated light curves, and Table~\ref{tab:tde_b_beta} gives the mapping between its impact-parameter scaling $b$ and the penetration factor $\beta_\mathrm{orb}$.

\begin{table}
    \centering
    \caption{Survey bands used in the analysis: effective frequency and wavelength at which the spectral energy distribution is evaluated from \texttt{Redback}~\citep{Sarin:2023uqa} and median revisit time per band, from the recorded pointing histories of \ac{ZTF}~\citep{Bellm:2018cps, Masci:2018neq} and the \ac{LSST}~\citep{Bianco:2021ape}.}
    \label{tab:bands_cadence}
    \begin{tabular}{llccc}
    \hline
    Survey & Band & $\nu_{\mathrm{eff}}$ ($10^{14}$~Hz) & $\lambda_{\mathrm{eff}}$ (\AA) & Median revisit (d) \\
    \hline
    \multirow{3}{*}{\ac{ZTF}} & $g$ & 6.272 & 4784 & 1--2 \\
     & $r$ & 4.675 & 6417 & 1--2 \\
     & $i$ & 3.813 & 7867 & 3--4 to $\gtrsim$15--25 \\
    \hline
    \multirow{6}{*}{\ac{LSST}} & $u$ & 8.152 & 3680 & 20--30 \\
     & $g$ & 6.273 & 4782 & 20--30 \\
     & $r$ & 4.825 & 6218 & 5--7 \\
     & $i$ & 3.983 & 7532 & 6--8 \\
     & $z$ & 3.454 & 8685 & 5--7 \\
     & $y$ & 3.083 & 9730 & 10--16 \\
    \hline
    \end{tabular}
\end{table}

\begin{table}
    \centering
    \caption{Parameters of the \texttt{tde\_fallback} model. Parameters labelled fixed take the stated value for all events~\citep{Lodato:2008fr, Mockler:2018xne}}
    \label{tab:tde_lc_priors}
    \begin{tabular}{llll}
    \hline
    Parameter & Symbol & Prior & Range or value \\
    \hline
    Stellar mass & $M_\star$ & Fixed & $1~M_\odot$ \\
    Impact-parameter scaling & $b$ & Uniform & $[0,2]$ \\
    Radiative efficiency & $\eta$ & Log-uniform & $[10^{-3},10^{-1}]$ \\
    Eddington fraction & $f_{\mathrm{Edd}}$ & Fixed & $1$ \\
    Viscous timescale & $t_{\mathrm{visc}}$ & Log-uniform & $[1,30]$~day \\
    Photosphere amplitude & $r_{\mathrm{ph},0}$ & Log-uniform & $[10^{-4},10^{4}]$ \\
    Photosphere index & $l_{\mathrm{photo}}$ & Uniform & $[0,2]$ \\
    Hydrogen fraction & $X_\mathrm{H}$ & Fixed & $0.74$ \\
    \hline
    \end{tabular}
\end{table}

\begin{table}
    \centering
    \caption{Relation between the \texttt{mosfit} impact-parameter scaling $b$ and the penetration factor $\beta_{\mathrm{orb}}=r_\star/r_{\mathrm{p}}$ for the two polytropic indices $\gamma$ of the fallback-rate simulations of \citet{Guillochon:2012uc} implemented in \texttt{mosfit}~\citep{Mockler:2018xne} and \texttt{Redback}~\citep{Sarin:2023uqa}.}
    \label{tab:tde_b_beta}
    \begin{tabular}{lcc}
    \hline
    $\gamma$ & $\beta_{\mathrm{orb}}(b)$, $0\le b<1$ & $\beta_{\mathrm{orb}}(b)$, $1\le b\le 2$ \\
    \hline
    $4/3$ & $0.6+1.25\,b$ & $1.85+2.15\,(b-1)$ \\
    $5/3$ & $0.5+0.4\,b$ & $0.9+1.6\,(b-1)$ \\
    \hline
    \end{tabular}
\end{table}

\bibliographystyle{mnras}
\bibliography{biblio}

\bsp	
\label{lastpage}
\end{document}